\documentclass{article}

\usepackage{PRIMEarxiv}

\usepackage[utf8]{inputenc} 
\usepackage[T1]{fontenc}    
\usepackage{hyperref}       
\usepackage{url}            
\usepackage{booktabs}       
\usepackage{amsfonts}       
\usepackage{nicefrac}       
\usepackage{microtype}      
\usepackage{lipsum}
\usepackage{fancyhdr}       
\usepackage{graphicx}       
\graphicspath{{media/}}     

\usepackage{multirow}%
\usepackage{amsmath,amssymb}%
\usepackage{amsthm}%
\usepackage{mathrsfs}%
\usepackage[title]{appendix}%
\usepackage{xcolor}%
\usepackage{textcomp}%
\usepackage{manyfoot}%
\usepackage{listings}%
\usepackage{caption}
\usepackage{rotating}

\title{Double-tough and ultra-strong ceramics: leveraging multiscale toughening mechanisms through Bayesian Optimization}

\author{Francesco Aiello\textsuperscript{1}\hspace{10pt}
Jian Zhang\textsuperscript{1}\hspace{10pt}
Johannes C. Brouwer\textsuperscript{2}\hspace{10pt}
Mauro Salazar\textsuperscript{1*}\hspace{10pt}
Diletta Giuntini\textsuperscript{1,3*} 
\medskip\\
\textsuperscript{1} Department of Mechanical Engineering, Eindhoven University of Technology,\\
P.O. Box 513, Eindhoven, 5600 MB, The Netherlands.\\
\textsuperscript{2} Department of Materials Science and Engineering, Delft University of Technology,\\
Mekelweg 2, Delft, 2628 CD, The Netherlands.\\
\textsuperscript{3} Institute of Advanced Ceramics, Hamburg University of Technology,\\
Denickestrasse 15, Hamburg, 21073, Germany.
\medskip\\
\textsuperscript{*}Corresponding authors: M. Salazar (\texttt{m.r.u.salazar@tue.nl}); D. Giuntini (\texttt{d.giuntini@tue.nl}).}

\begin{document}
\maketitle

\begin{abstract}
An optimization-driven approach is presented to create a "double-tough" ceramic material with a brick-and-mortar microstructure, where the mortar is itself transformation-toughened, engineered to achieve high strength and fracture toughness levels simultaneously. The material design involves high-strength alumina bricks interconnected via a ceria-stabilized zirconia mortar. Given that the design, driven by multiscale toughening mechanisms, typically requires a laborious trial-and-error approach, a Bayesian optimization framework is proposed to streamline and accelerate the design process. A Gaussian process is used to emulate the material's mechanical response, and a cost-aware batch Bayesian optimization is implemented to efficiently identify optimal design process parameters, accounting for the cost of experimentally varying them. This approach expedites the optimization of the material's mechanical properties. As a result, a bio-inspired all-ceramic composite is developed, exhibiting an exceptional balance between bending strength ($704\,$MPa) and fracture toughness ($13.6\,$MPa$\,$m$^{0.5}$).
\end{abstract}

\keywords{Ceramics \and Bio-inspired materials \and Transformation toughening \and Spark plasma sintering \and Field-assisted sintering \and Sol-gel process \and Bayesian Optimization.}

\section{Introduction}
\label{section:introduction}
Ceramic materials are renowned for their exceptional hardness and strength, coupled with high-temperature and chemical stability. However, their inherent brittleness presents significant challenges, limiting their broad applicability~\cite{rosler2007mechanical}. Balancing strength and toughness is a common challenge in materials science, as they are two crucial yet conflicting properties~\cite{ritchie2011conflicts}.	A well-established solution relies on the phase transformation toughening featured by zirconia, whilst more recently bio-inspired nacre-like structures have started to prove effective in leading to tough ceramic-based composites ~\cite{huang2019multiscale,green2018transformation,plunkett2022bridging,bor2020mapping,magrini2019transparent,magrini2020transparent,grossman2017mineral,poloni2021tough,magrini2022fracture,chan2022energy}. These different toughening mechanisms, however, are often thought to be incompatible, even though combining both in one and the same material could lead to remarkable steps forward in the quest of strong and tough all-ceramic materials. 

We fabricate here an all-ceramic composite that exploits these two primary toughening mechanisms: crack deflection, through a brick-and-mortar microstructure, and crack shielding, via transformation toughening within the mortar, yielding a "double-tough" ceramic composite. The crack deflection mechanism operates at the microscale (the scale of the bricks), the crack shielding at the nanoscale (in the sub-$\mu$m-thick mortar). 

There are several challenges in developing such a double-tough ceramic: (1) The phase transformation in zirconia requires a minimum grain size to operate at realistic stresses, whilst nacre-mimetics typically need ultrathin mortars; (2) the parameter space to explore, in terms of material composition, microstructure and process parameters, is extremely broad. To overcome these challenges and achieve the goal of maximizing both strength and toughness, we leverage Bayesian optimization methods to expedite the experimental campaign.

This paper thus touches upon three research streams: nacre-like ceramics, stress-induced phase-transforming materials, and Bayesian optimization for design purposes.
Nacre-like composites aim to replicate the remarkable hierarchical structure of nacre, known for its exceptional combination of strength and toughness~\cite{zhao2017nacre}. This superior toughness in nacre arises from various mechanisms inherent to its multi-scale hierarchical architecture~\cite{sun2012hierarchical,grossman2018quantifying}, with the "brick-and-mortar" microstructure playing a central role~\cite{grossman2019hierarchical,wang2001deformation,radi2021effect,saad2022toughening}.
Platelets of aluminum oxide (Al$_2$O$_3$) exhibit high micromechanical flexural strength~\cite{feilden2017micromechanical} and are commonly used in constructing nacre-like structures~\cite{corni2012review,le2023nacre,le2019processing}. In this context, nacre-like alumina, densified through spark plasma sintering, exhibited enhanced strength, particularly under dynamic loading conditions~\cite{evers2020nacre}. Nevertheless, it is widely recognized that a ductile mortar phase serves a critical role in composite materials, contributing to various functions, such as facilitating crack deflection~\cite{clegg1990simple}, distributing stress around crack tips~\cite{yeom2006effect}, and dissipating energy through plastic deformation~\cite{bonderer2010platelet,niebel2016role}.
Leon et al.~\cite{dimas2013tough} employed a mesoscale bead model to illustrate how materials with an elastic behavior until the point of fracture can be strategically used to fashion "brick-and-mortar" composites with remarkable stiffness and superior toughness. Their research emphasized how the compliant mortar nature serves to delocalize stresses and strains away from the crack tip, resulting in substantial crack deflection.
In the context of nacre-like alumina, various secondary phases were explored to enhance the material's properties, even though mostly non-ceramic ones. Remarkable exceptions are found in the incorporation of secondary phases such as silica and calcia~\cite{bouville2014strong} or aluminum borate~\cite{pelissari2018nacre,pelissari2020refractory}, which allow for a favorable balance between toughness and strength. Furthermore, the ceramic mortar can maintain high temperatures and corrosion resistance.	Based on these observations, zirconium oxide, often known as zirconia (ZrO$_2$), emerges as a promising candidate for the "mortar" phase in nacre-like all-ceramic composites. Zirconia is more compliant than alumina~\cite{pabst2004effective} and possesses remarkable toughness due to the stress-induced tetragonal to monoclinic phase transformation~\cite{green2018transformation}.
The toughening mechanism takes place when tetragonal zirconia (t-ZrO$_2$) transforms into the monoclinic phase (m-ZrO$_2$) due to the application of external stress. When the transformation is initiated by stress concentrations, such as those caused by a crack, the resulting expansion leads to the formation of a beneficial compressive stress field around the crack tip. This phenomenon, known as crack tip shielding, significantly enhances the material's toughness~\cite{green2018transformation}. 

The retention of tetragonal zirconia is thus crucial. In equilibrium conditions, large particles of pure tetragonal zirconia initiate their transformation into the monoclinic phase during cooling at approximately $1150\,^\circ$C~\cite{hannink2000transformation}. Nonetheless, it is well-documented that if the grain size is smaller than a critical value, this transformation can occur at lower temperatures, thus enabling the retention of the tetragonal phase at the material's operating temperature~\cite{garvie1985part1}. The stress-induced transformation can toughen the material if the grain size falls within a specific range. Larger grains generate a spontaneous tetragonal to monoclinic transformation, and smaller grains make the tetragonal to monoclinic transformation thermodynamically unfavorable, resulting in a stable tetragonal phase even under high applied stresses~\cite{garvie1985part2}. Retention of the tetragonal phase can be controlled by introducing solid solutions with stabilizing oxides, such as yttria (Y$_2$O$_3$) and ceria (CeO$_2$), or by adjusting the grain size~\cite{green2018transformation}. Existing literature indicates that a grain size close to the critical value for spontaneous transformation maximizes toughness in nanocrystalline zirconia~\cite{bravo2002fracture}. A zirconia mortar then needs to be sized compatibly with the critical grain size to achieve transformation toughening. However, in  nacre-like ceramics the material's strength is observed to rise as the interphase thickness decreases~\cite{budarapu2021micromechanics}. These dimensional requirements create a conflict between the two toughening mechanisms, necessitating the identification of an optimal value. 

Traditionally, material optimization involves the use of a trial-and-error approach. Initially, the material is created using existing theoretical knowledge. Subsequently, it is iteratively refined by drawing upon the results of testing and fostering semi-empirical relationships that connect process parameters with material properties, all with the aim of enhancing previous outcomes. This conventional procedure is often time-consuming, reliant on approximations that are imprecise and only partially applicable, and it frequently overlooks the complex interactions among multiple variables. The advancement in computational capabilities and the accessibility of advanced optimization methods has opened avenues for expediting the design process. By maintaining a trial-and-error approach while enhancing decision-making based on insights from test results, we can cleverly select the next combination of design parameters, leading to a more efficient material optimization process.
Bayesian Optimization (BO) is a technique specifically tailored to optimize objectives that are computationally expensive, subject to noisy values, lack a straightforward analytical formulation, and present challenges in efficiently determining gradients. For a more comprehensive understanding of BO, the reader can consult the following textbook~\cite{garnett2023bayesian}. BO employs a probabilistic surrogate model, typically based on a Gaussian process~\cite{garnett2023bayesian,williams2006gaussian}, to estimate how the objective function behaves at unexplored locations, and an acquisition function to determine the next sampling point. In BO, the simultaneous evaluation of multiple points has the potential to expedite the optimization process~\cite{greenhill2020bayesian}. Sequential algorithms allow to choose a batch of points by selecting the one that maximizes the acquisition function before moving on to locate the next point within the batch~\cite{gonzalez2016batch}. BO can thus reduce the number of evaluations needed to discover optimal solutions~\cite{diessner2022investigating}, aligning with the requirements of material design optimization~\cite{frazier2015bayesian}. Its successful applications span various domains, from streamlining the synthesis of short polymer fibers~\cite{li2017rapid} to enhancing extracellular vesicle production~\cite{bader2023improving}, optimizing electric machines~\cite{borsboom2023effective} and biopharmaceutical formulations~\cite{narayanan2021design}, and integrating with numerical models in the field of material design~\cite{tran2020multi,balachandran2016adaptive,yamawaki2018multifunctional,deshwal2021bayesian,park2023multi}. Recently, BO explored the optimal balance between strength, toughness, and specific volume within a numeric finite element model of a nacre-inspired composite~\cite{park2023multi}. The study investigated a mortar with compliance ranging from one to three orders of magnitude higher than that of the bricks, consistent with a polymer-ceramic nacre-inspired composite.

To the best of the authors' knowledge, the integration of a "brick and mortar" structure with a material featuring transformation toughening remains an unexplored frontier due to challenges that BO can help tackling. 

In this study, BO enables the development of materials that leverage two toughening mechanisms with contrasting dimensional requirements, which occur simultaneously across different length scales. The result is an all-ceramic composite material featuring a "brick-and-mortar" microstructure, where alumina "bricks" are interconnected through a zirconia "mortar" that can undergo a phase transformation. To enhance the toughness and low-temperature degradation resistance, we employ ceria to stabilize the zirconia in its tetragonal phase~\cite{grathwohl1991crack,kohorst2012low,inserra2023preparation}. The material processing involves a sol-gel method to coat alumina platelets with nanograined ceria-stabilized zirconia~\cite{shen2010synthesis,rossignol1999preparation}, followed by spark plasma sintering of the coated platelets. We investigate the impact of process parameters on the strength and toughness of the material, ultimately identifying the optimal conditions. To streamline this optimization process, we employ Bayesian optimization (BO) and integrate a penalized acquisition function to efficiently assemble batches of points, taking into account the cost associated with evaluating different parameters within a single batch~\cite{garnett2023bayesian,gonzalez2016batch}.

\section{Meta-experimental Optimization Methodology}
We fabricate the material by coating commercial Al$_2$O$_3$ platelets with ceria-stabilized zirconia, and then sinter the resulting powder using spark plasma sintering (see Experimental Section). The coating procedure allows us to obtain a sub-µm zirconia layer, which, once sintered, becomes confined between the alumina platelets, making the mortar phase in the "brick-and-mortar" composites.

We evaluate the effectiveness of the platelet coating and examine the microstructure of the final material through the use of a scanning electron microscope equipped with an energy-dispersive x-ray analyzer. Additionally, we determine crystal structures of the different phases through x-ray diffractometric analysis, confirming the presence of tetragonal zirconia, which is crucial for exploiting transformation toughening via stress-induced tetragonal to monoclinic phase transformation. We determine the mechanical properties of the sintered material through three-point bending tests to measure bending strength and crack growth resistance curve (R-curve). We quantify toughness by calculating $K_{J,0}$ at crack initiation and $K_{J,\mathrm{lim}}$ at the crack extension limit, as defined by the standard ISO 12135:2021~\cite{iso12135}.

The process parameters involved are numerous and can take on continuous values, including the calcination and sintering temperatures, pH of the precursor solution, and platelet aspect ratio. Therefore, discretizing and testing the multiple possible combinations of process parameters results in the testing of a vast number of experiments, which increases exponentially with the number of parameters considered, rendering this method impractical.

To obviate this problem, we first reduce the domain size by selecting the process parameters that significantly affect mechanical properties and thus merit initial investigation. Our priority is set on the ceria mole fraction, $x_{\mathrm{CeO}_2}$, mortar mass ratio, $X_{\mathrm{m}}$, and dwell temperature, $T_{\mathrm{dwell}}$. Specifically, the amount of mortar impacts stress redistribution and defect tolerance~\cite{budarapu2021micromechanics}, the ceria content affects the chemical free energy of the tetragonal to monoclinic phase transformation and resultant toughening~\cite{garvie1985part1,garvie1985part2}, and furthermore, dwell temperature affects material densification and grain growth~\cite{castro2012sintering}.

Secondly, we formulate an objective function, $f(\mathbf{x})$, as a linear combination of normalized flexural strength and stress intensity factor at the limit, where $\mathbf{x}$ represents the selected process parameters. We employ BO to maximize this function. In contrast to traditional design of experimental methods, which typically involve fixed experiment numbers and limited adaptability, BO follows a sequential approach~\cite{narayanan2021design}. It relies on a surrogate model of the actual system and a strategy to determine the subsequent experiment based on existing data.

The surrogate model is here based on a Gaussian Process (GP). We leverage our knowledge of the function $f(\mathbf{x})$ to train the GP and make predictions for its expected value at a given test point.
To initialize the BO, we train the GP by selecting an initial set of process parameters based on literature observations. For instance, the mortar content in natural nacre typically ranges around $5\,$vol$\%$~\cite{sun2012hierarchical}, while in brick-and-mortar-like materials, it can increase significantly, reaching up to $45\,$vol$\%$~\cite{bouville2020strong,budarapu2021micromechanics}. In zirconia-toughened alumina, the zirconia content is generally kept below $30\,$wt.$\%$~\cite{sommer2012mechanical,de2002crack,zadorozhnaya2020effect,pfeifer2016synthesis} due to challenges in maintaining dispersed zirconia grains in the tetragonal phase above $16\,$vol$\%$~\cite{sommer2012mechanical,lange1982transformation4}. Ceria serves as a stabilizing agent for tetragonal zirconia at room temperature~\cite{nettleship1987tetragonal}. The solubility limit of the tetragonal solid solution is around $18\,$mol$\%$ ceria at $1400\,^\circ$C, decreasing to $16\,$mol$\%$ at $1200\,^\circ$C~\cite{tani1983revised}. Typically, ceria content in ceria-stabilized zirconia is less than $12\,$mol$\%$~\cite{quinelato2000effect,rezaei2009synthesis,grathwohl1991crack,liu1994effects,liens2020phase,fornabaio2014zirconia,sharma2002synthesis}. Higher ceria concentrations result in the formation of a stable cubic zirconia phase~\cite{nettleship1987tetragonal,tani1983revised,sharma2002synthesis}. The stability of tetragonal zirconia depends on grain size~\cite{garvie1985part1}, leading to the concept of critical size for spontaneous transformation to the monoclinic phase. Dwell temperature selection is crucial to achieve final density without excessive grain growth or induced monoclinic transformation during sintering~\cite{orlovska2022study,nettleship1987tetragonal}.

The objective of the acquisition strategy is to select exploratory points to investigate, through a balance between exploring new areas of the objective function and exploiting known information to improve the current best point~\cite{garnett2023bayesian}.
We adopt a sequential algorithm introduced by Gonz{\'a}lez et al.~\cite{gonzalez2016batch} to obtain a batch of exploration points. We modify this algorithm to consider the cost associated with exploring different combinations of process parameters within a single batch. Since adjustments to the ceria content or mass ratio of the mortar necessitate the creation of distinct materials, while modifications to the dwell temperature can be easily applied to two samples of material with identical compositions, subsequent batch points are deemed cost-effective for evaluation if the newly selected process parameters do not alter the ceria content or mass ratio of the mortar.

\section{Pre-optimization Material Processing}
\label{res:pre-optimization}
\begin{figure}
	\includegraphics[width=\textwidth]{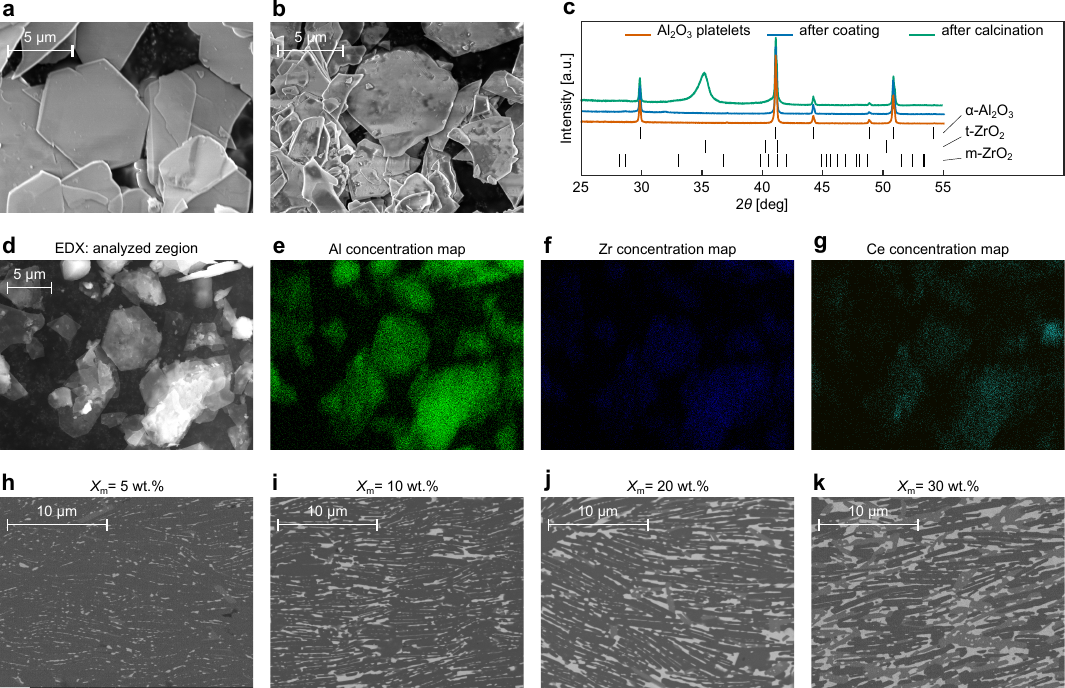}
	\caption{Pre-optimization analysis. (a, b) SEM images of alumina platelets before and after being coated with $10\,$wt.$\%$ ceria-doped zirconia mortar, showing effective material coverage. (c) XRD spectra of Al$_2$O$_3$ platelets, right after being coated with $30\,$wt.$\%$ mortar and $12\,$mol$\%$ ceria, and after being calcined, revealing a fully tetragonal zirconia coating. Theoretical diffraction peaks for alpha alumina ($\alpha$-Al$_2$O$_3$), tetragonal zirconia (t-ZrO$_2$), and monoclinic zirconia (m-ZrO$_2$) are shown. (d-g) SEM images with EDX analysis of alumina platelets coated with $20\,$wt.$\%$ zirconia stabilized with $6\,$mol$\%$ ceria conducted under high vacuum conditions with a beam voltage of $10\,$kV, showcasing elemental concentration with homogeneous distribution. (h-k) SEM images of material microstructures with varying mortar content $(5$-$30\,$wt.$\%$), with zirconia as the light phase and alumina as the dark phase. The microstructures reveal the alignment of alumina platelets in a preferential direction, with the mortar becoming continuous and more uniform as its content increases. Images (a, b, d-g) obtained under low vacuum conditions, with a chamber pressure of $0.6\,$mbar, working distance of $5\,$mm, beam voltage of $5\,$kV, and spot size set to $3$; micrographs (d-g) obtained using a back-scattered electron detector in a plane aligned with the sintering direction. }
	\label{fig:pre_optimization_analysis}
\end{figure}
In the pre-optimization phase, we analyzed the coating uniformity and the zirconia phases present in the prepared powder. Figure~\ref{fig:pre_optimization_analysis}~(a, b) presents comparative SEM images of alumina platelets in their as-received state (Figure~\ref{fig:pre_optimization_analysis}~(a)) and after undergoing the coating and deagglomeration process (Figure~\ref{fig:pre_optimization_analysis}~(b)), before the subsequent calcination. Figure~\ref{fig:pre_optimization_analysis}~(b) illustrates representative coated platelets, demonstrating qualitative evidence of material coverage without observable agglomerations.

Figure~\ref{fig:pre_optimization_analysis}~(c) depicts XRD spectra obtained from alumina platelets in their as-received state, after coating (pre-calcination), and immediately after calcination.
The as-received platelets exhibited peaks corresponding to the theoretical peaks of $\alpha$-Al$_2$O$_3$. Post-coating, no other crystalline phases were evident, as indicated by the absence of additional peaks in the spectrum. Upon calcination, a crystalline phase emerged, notably characterized by a peak at around $35\,$degrees, corresponding to the diffraction peak of t-ZrO$_2$. No peaks indicative of m-ZrO$_2$ were observed in the spectrum. After refinement, we determined the crystallite size of tetragonal zirconia to be $13\,$nm, consistent with the grain size of the zirconia coating obtained using a similar method~\cite{shen2010synthesis}.

We conducted EDX analysis on the coated platelets after calcination to assess the uniformity of elements. Figure~\ref{fig:pre_optimization_analysis}~(d) displays a SEM image of alumina platelets coated with $20\,$wt.$\%$ of mortar and $6\,$mol$\%$ of ceria, acquired at a $10\,$kV beam voltage, with corresponding aluminum, zirconium, and cerium maps shown in Figure~\ref{fig:pre_optimization_analysis}~(e-g).
The elements exhibited homogeneous distribution. The low intensity of the cerium element map is attributed to its content, constituting $6\,$mol$\%$ of the mortar phase, which, in turn, represented $20\,$wt.$\%$ of the material. A detailed elemental scan revealed a zirconium-to-cerium ratio of $9.45$:$0.65$, consistent with the theoretical $6\,$mol$\%$ present in the analyzed coating.
We conduct XRD analysis to monitor zirconia phases in the coating before and after calcination. 

The microstructures resulting from sintering are depicted in Figure~\ref{fig:pre_optimization_analysis}~(h-k), illustrating microstructural images for varying amounts of mortar, ranging from $5\,$wt.$\%$ to $30\,$wt.$\%$. We captured the images along a plane aligned with the direction of load application during sintering, oriented vertically in the figures. We used back-scattered electron detector imaging to emphasize the contrast between the "brick" and "mortar" phases, corresponding to the dark and light phases, respectively, due to the distinct atomic numbers of Al and Zr~\cite{hawkes2007science}. The microstructures revealed the alignment of alumina platelets in a preferential direction. The orientation, orthogonal to the midplane (parallel to the crystallographic $c$-axis), was aligned parallel to the direction of the applied load during sintering. Figure~\ref{fig:pre_optimization_analysis}~(h), shows the mortar phase discontinuously distributed around the platelets, with absence between some platelets and small agglomerates observed on the sides of the alumina platelets. As the mortar content increased, the thickness of the mortar phase became more continuous and uniform, as depicted in Figure~\ref{fig:pre_optimization_analysis}~(i-k). 

In the following three sections, we explore findings from specimens made during the optimization phase, examining how the process parameters affected the final material. To streamline the naming convention for the tested compositions, we used a representation that indicates both the mortar content in wt.$\%$ and the ceria concentration in mol$\%$. For instance, "$10X_{\mathrm{m}}$-$12x_{\mathrm{CeO}_2}$" denoted a composition with the $10\,$wt.$\%$ in mortar mass ratio and $12\,$mol$\%$ in CeO$_2$ mole percentage.

\section{Material Optimization}
\begin{figure}
	\centering
	\includegraphics[width=\textwidth]{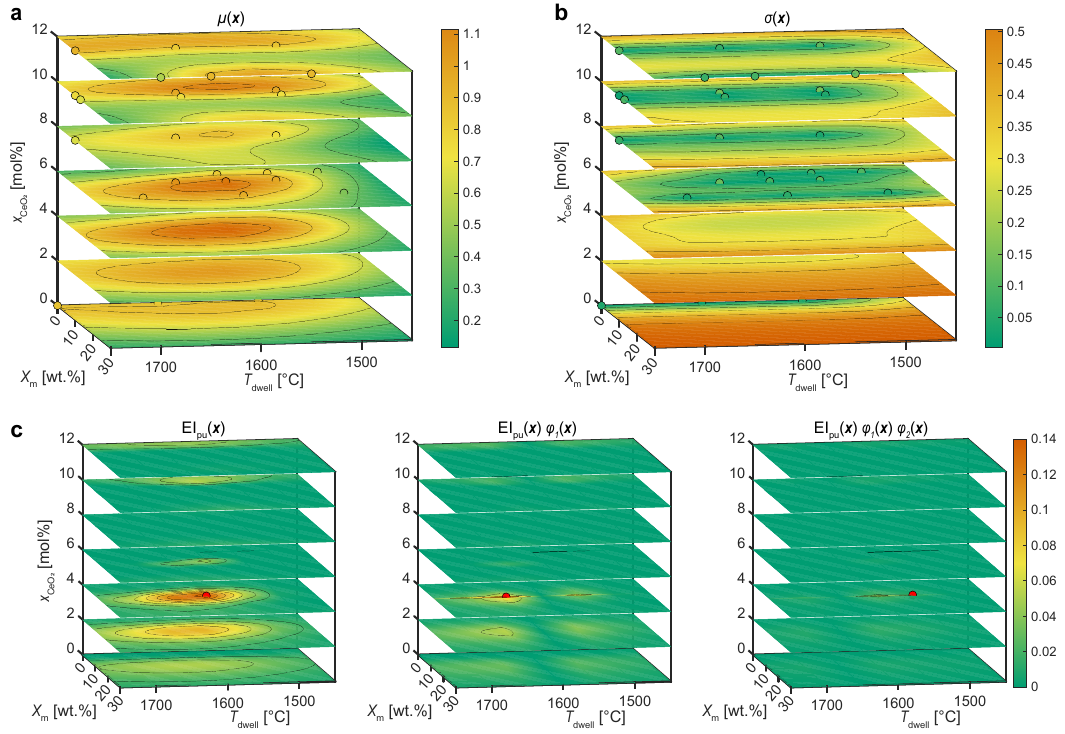}
	\caption{Illustration of Gaussian Process and exploration points determination. Visualization of the Gaussian Process (GP) representing (a) the posterior mean ($\mu(\mathbf{x})$) and (b) the posterior standard deviation ($\sigma(\mathbf{x})$) of the objective function $f(\mathbf{x})$ with the associated training data set. The mean function relates process parameters to the expected value of the objective function, while the standard deviation provides information on uncertainty, with higher values indicating greater uncertainty. In (c), acquisition functions across three iterations are shown, aiming to select a batch of three points for exploration. The coordinates of the maximum point in each acquisition function, denoted by a red dot, represent the process parameters for the batch points. Progressing from left to right, the Expected Improvement per unit of cost ($\mathrm{EI_{pu}}$) is adjusted with the current exploration cost, and the point selected in the previous iteration is penalized by the function $\varphi$ in the subsequent iteration. Functions are expressed in terms of mortar content ($X_{\mathrm{m}}$), dwell temperature ($T_{\mathrm{dwell}}$), and ceria mole fraction ($x_{\mathrm{CeO}_2}$), depicted through multiple contour plots at constant $x_{\mathrm{CeO}_2}$ values. In (a, b), the function is overlaid with a scatter plot representing the mean and standard deviation of the dataset.}
	\label{fig:gp_af}
\end{figure}

Our objective in material optimization was to determine the optimal values for ceria content $x_{\mathrm{CeO}_2}$, mortar amount $X_{\mathrm{m}}$, and dwell temperature $T_{\mathrm{dwell}}$ that maximize the objective function defined in equation~\eqref{eq:optimization_problem} (Experimental Section). We selected reference parameters $\bar{K}_{J,\mathrm{lim}}$ and $\bar{\sigma}_{\mathrm{f}}$ with values of $16.9\,\mathrm{MPam^{0.5}}$ and $450\,\mathrm{MPa}$, respectively. These values represented a well-balanced combination of mechanical properties for nacre-like alumina~\cite{bouville2020strong}. We set the weight parameter $\lambda$ in equation~\eqref{eq:optimization_problem} (Experimental Section) to $0.5$ to achieve a balance between maximizing strength ($\lambda=0$) and toughness ($\lambda=1$). Additionally, we fixed the batch size at $3$. The values for ceria concentration, mortar content, and dwell temperature were rounded to the nearest multiples of $1\,$mol$\%$, $1\,$wt.$\%$, and $50\,^\circ$C, respectively.	
Before initiating the optimization via BO, based on the considerations made in the 'Material Optimization' part of the 'Experimental Section', we selected an initial set of compositions to train the GP. We chose mortar mass fraction, $X_{\mathrm{m}}$, ranging from $0$ (alumina platelets only) to $30\,$wt.$\%$, ceria mole fraction, $x_{\mathrm{CeO}_2}$, up to $12\,$mol$\%$, and dwell temperatures, $T_{\mathrm{dwell}}$, below $1750\,^\circ$C. The training points were then recursively expanded with exploration points suggested by the acquisition strategy, with the goal of solving the optimization problem posed in equation~\eqref{eq:optimization_problem} (Experimental Section).	As an illustrative example of the procedure adopted, Figure~\ref{fig:gp_af} illustrates the GP during the definition of the last batch of explored points. Figure~\ref{fig:gp_af}~(a, b) displays the GP posterior mean $\mu(\mathbf{x})$, and the GP posterior standard deviation $\sigma(\mathbf{x})$, alongside the mean and standard deviation of the training data set, respectively. The GP posterior mean revealed two local maxima at ceria concentrations of $4$ and $10\,$mol$\%$ in orange-yellow regions (Figure~\ref{fig:gp_af}~(a)). The GP posterior means calculated at these maxima exhibited different estimated dispersions, as indicated by the respective standard deviations (Figure~\ref{fig:gp_af}~(b)). Notably, the maximum at $4\,$mol$\%$ of ceria was situated in a relatively unexplored region, leading to a greater uncertainty in its estimated value.	The acquisition function used $\mu(\mathbf{x})$ and $\sigma(\mathbf{x})$ to identify exploration points. Figure~\ref{fig:gp_af}~(c) illustrates the values assumed by the acquisition function over three iterations for batch point exploration. In the left figure, the acquisition function is represented by the expected improvement per unit of cost $\mathrm{EI_{pu}}$, which coincides with EI since the cost of the first exploration point is unitary. The function peaks at $[13\,\mathrm{wt.}\%, 4\,\mathrm{mol}\%, 1600\,^\circ\mathrm{C}]$, identifying the initial point of the batch $\mathbf{x}^b_1$. This selected point displayed a high a posteriori estimated value for the objective function and a substantial standard deviation, making it a favorable candidate for both exploration and exploitation strategies~\cite{garnett2023bayesian}. In the following figure, the selection of $\mathbf{x}^b_1$ was penalized by the penalty function $\varphi_1$. Consequently, the cost function is updated to prioritize the choice of an exploration point with the same composition, considering it more cost-effective. The maximization of the updated acquisition function leads to the selection of the next exploration point ($\mathbf{x}^b_2 = [13\,\mathrm{wt.}\%, 4\,\mathrm{mol}\%, 1650\,^\circ\mathrm{C}]$), and this process is repeated for the third point ($\mathbf{x}^b_3 = [13\,\mathrm{wt.}\%, 4\,\mathrm{mol}\%, 1550\,^\circ\mathrm{C}]$).
\begin{figure}
	\centering
	\includegraphics[width=\textwidth]{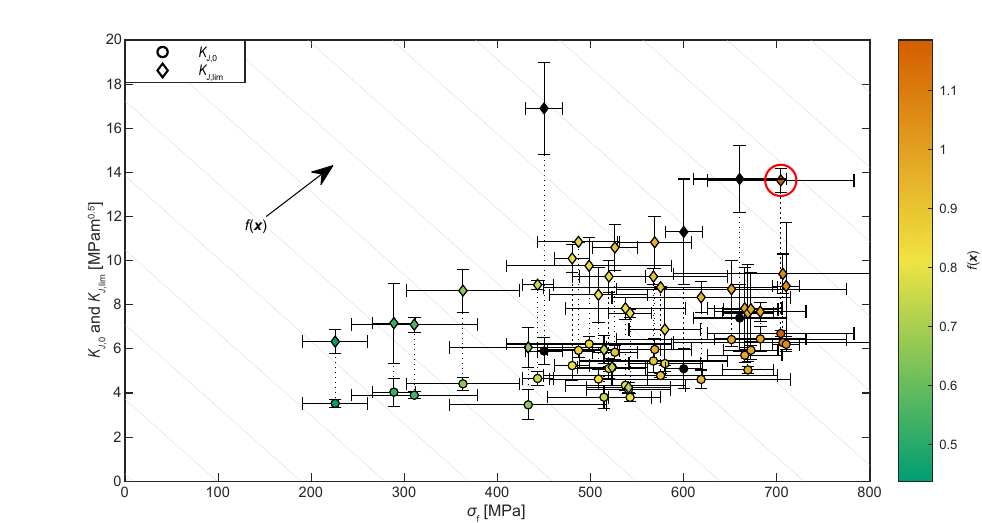}
	\caption{Comparison of mechanical properties. Map of flexural strength ($\sigma_{\mathrm{f}}$) versus stress intensity factor at both crack initiation ($K_{J,\mathrm{0}}$) and limit ($K_{J,\mathrm{lim}}$). Black markers represent literature data for nacre-like composites consisting solely of ceramic components~\cite{bouville2014strong,le2015magnetically,pelissari2018nacre}, while colored markers indicate data points from the current study, with colors corresponding to the objective function value according to the color bar. A red circle highlights the tested material with the highest objective function value. Light blue lines indicate regions with a constant objective function value, with an arrow indicating the direction of objective function growth. Error bars represent one standard deviation.}
	\label{fig:data_comparison}
\end{figure}

The mechanical properties and corresponding values of the objective function for all sampled process parameter combinations are reported in Table~S2 in the Supporting Information. While the optimization procedure could have continued to identify additional potential candidates as maximum points, we chose to interrupt the process due to the updated maximum EI, which was equal to $6\%$. Figure~\ref{fig:data_comparison} summarizes some of the explored points alongside literature data obtained on nacre-like alumina, represented in a diagram of flexural strength versus stress intensity factor at both crack initiation and limit. The maximum point (coordinates: $[10\,\mathrm{wt.}\%, 6\,\mathrm{mol}\%, 1600\,^\circ\mathrm{C}]$) achieved a mean value of flexural strength and stress intensity factor at the limit equal to $704\,\mathrm{MPa}$ and $13.6\,\mathrm{MPam^{0.5}}$, respectively. It exhibited superior strength compared to existing materials while still maintaining very good fracture toughness.

\section{Influence of Process Parameters on Phase Composition, Microstructure and Density}
\label{sec:InfluenceOfProcessParametersOnPhaseComposition}
The x-ray diffraction analysis indicated the presence of $\alpha$-Al$_2$O$_3$, t-ZrO$_2$, and m-ZrO$_2$ phases exclusively. By refining the spectra, we determined the quantities of each phase and their respective crystallite sizes. For a visual representation of the XRD spectra and the refinement outcomes, please see Figure~S1 and Table~S1 in the Supporting Information. The alumina content, as determined through XRD spectra refinement, exhibited a deviation from the nominal value, with a maximum difference of approximately $5\%$.
\begin{figure}
	\centering
	\includegraphics[width=\textwidth]{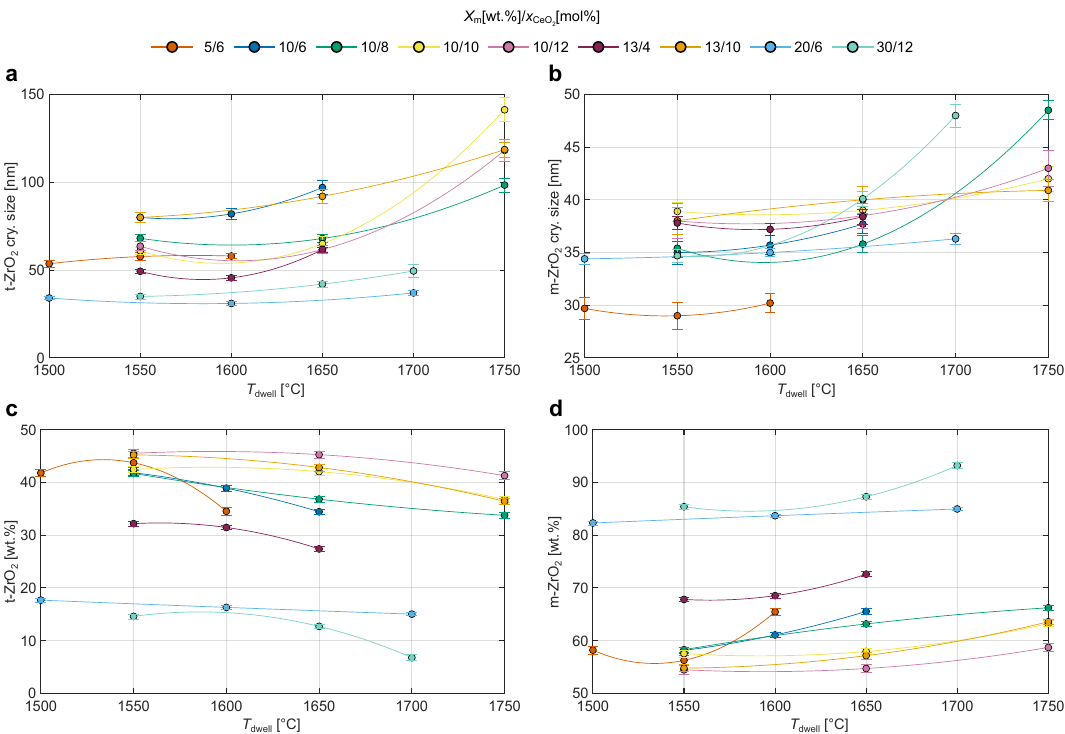}
	\caption{Variation in crystallite size and phase content of zirconia mortar in tested compositions. Variation in the crystallite size of the t-ZrO$_2$ phase (a) and the m-ZrO$_2$ phase (b), and percentage by weight of the t-ZrO$_2$ phase (c) and the m-ZrO$2$ phase (d) in the mortar, as a function of the dwell temperature ($T_{\mathrm{dwell}}$) for different tested combinations of mortar mass ratio ($X_{\mathrm{m}}$) and ceria mole fraction ($x_{\mathrm{CeO}_2}$). Error bars represent one standard deviation.}
	\label{fig:xrd_results}
\end{figure}
For a clearer visualization of the results, we depicted in Figure~\ref{fig:xrd_results} the crystallite sizes of both tetragonal and monoclinic zirconia, as well as the weight percentage of both tetragonal and monoclinic zirconia in the mortar for the various compositions as functions of $T_{\mathrm{dwell}}$. 

The size of tetragonal zirconia crystals increased with the rise in $T_{\mathrm{dwell}}$ across all tested compositions (Figure~\ref{fig:xrd_results}~(a)). Theoretically, a critical grain size for the tetragonal to monoclinic phase transformation in zirconia should impose an upper limit on the size of tetragonal crystals, as larger ones are expected to transform into the monoclinic phase. However, previous studies have consistently shown that tetragonal crystals grow monotonically rather than exhibiting a size limit~\cite{lin1998crystallite,shi1991effect}. This suggests that the tetragonal to monoclinic phase transformation cannot be solely attributed to the presence of a critical size, as other factors also play a role~\cite{lin1998crystallite,srinivasan1990critical}. In our "brick and mortar" structured material, wherein zirconia grains were confined between alumina platelets, residual stresses arising from the thermal expansion mismatch between alumina and zirconia played a role in stabilizing tetragonal zirconia~\cite{lange1982transformation1,becher1986toughening}. 

In Figure~\ref{fig:xrd_results}~(b), the crystallite size of the monoclinic phase is illustrated. Across all compositions, it exhibited an increase with the rise in $T_{\mathrm{dwell}}$, except for the $5X_{\mathrm{m}}$-$6x_{\mathrm{CeO}_2}$ specimen, where it remained relatively constant. The monoclinic zirconia crystal emerges through a phase transformation, dividing the preceding tetragonal grain into multiple segments~\cite{zhu1994x}. Consequently, the size of the resulting monoclinic zirconia crystals is reduced compared to the original tetragonal phase~\cite{zhu1994x}. This phenomenon is evident in the comparison between Figure~\ref{fig:xrd_results}~(a, b), showing that the crystallite size of the monoclinic phase was either comparable (as seen in specimen $20X_{\mathrm{m}}$-$6x_{\mathrm{CeO}_2}$) or smaller than that of the tetragonal phase. The growth rate of tetragonal crystals with respect to $T_{\mathrm{dwell}}$ exceeded that of the monoclinic phase. This difference is attributed to a higher oxygen lattice diffusion coefficient in the tetragonal phase, as highlighted by Ikuma et al.~\cite{ikuma1991oxygen}. Moreover, in Figure~\ref{fig:xrd_results}~(a), it was observed that at $T_{\mathrm{dwell}}$ between $1550$ and $1650\,^\circ$C, compositions with the same mortar amount (specimens $10X_{\mathrm{m}}$-$6x_{\mathrm{CeO}_2}$, $10X_{\mathrm{m}}$-$8x_{\mathrm{CeO}_2}$, $10X_{\mathrm{m}}$-$10x_{\mathrm{CeO}_2}$, and $10X_{\mathrm{m}}$-$12x_{\mathrm{CeO}_2}$) demonstrated that the increase in ceria content lead to a decrease in the size of tetragonal zirconia crystals. However, at higher $T_{\mathrm{dwell}}$, the crystal size appeared unrelated to ceria content. 

Figure~\ref{fig:xrd_results}~(c) illustrates the weight percentage of retained tetragonal zirconia in the mortar as a function of $T_{\mathrm{dwell}}$ for the various compositions. The remaining percentage consisted of monoclinic zirconia formed from the tetragonal phase during sintering. The retained amount of tetragonal zirconia in the mortar exhibited a decreasing trend as $T_{\mathrm{dwell}}$ increased. This phenomenon is well-documented, as an increase in $T_{\mathrm{dwell}}$ promotes the growth of zirconia crystals, facilitating martensitic transformation~\cite{becher1992grain,heuer1982stability}. This trend is influenced by both the amount of mortar and ceria content. In compositions featuring $10\,$wt.$\%$ mortar ($10X_{\mathrm{m}}$-$6x_{\mathrm{CeO}_2}$, $10X_{\mathrm{m}}$-$8x_{\mathrm{CeO}_2}$, $10X_{\mathrm{m}}$-$10x_{\mathrm{CeO}_2}$, and $10X_{\mathrm{m}}$-$12x_{\mathrm{CeO}_2}$), the retained tetragonal zirconia phase increased as ceria content rose for all $T_{\mathrm{dwell}}$ tested.
Notably, there was a substantial reduction in the retained tetragonal zirconia amount with higher mortar contents. For mortar contents exceeding $13\,$wt.$\%$, a sharp decline was observed, reaching levels of tetragonal zirconia in the mortar lower than $10\,$wt.$\%$ in the $30\,$wt.$\%$ mortar contents. The reduction in retained tetragonal zirconia with increasing mortar mass ratio can be attributed to autotransformation induced by tensile residual stress, as explained by Becher et al.'s theory~\cite{becher1986toughening}. According to their analytical model, zirconia particles embedded in an alumina matrix experience additional tensile residual stress transmitted through the matrix from neighboring zirconia particles. This reduces the external stress required to induce tetragonal to monoclinic transformation. The intensity of this additional residual stress is greater when zirconia particles are closer, as it decays with distance from each particle to the power of three. This phenomenon results in autotransformation of tetragonal zirconia particles when a critical amount of total transformable zirconia is reached. 

The maximum amount of retained tetragonal zirconia in the tested compositions was less than $50\,$wt.$\%$ of the mortar, and it varies depending on the sintering $T_{\mathrm{dwell}}$. Lin et al.'s study~\cite{lin1998crystallite} investigates the impact of aging temperature on the retention of tetragonal zirconia in ceria-stabilized zirconia. They observed minimal transformation of tetragonal zirconia in samples containing at least $5.5\,$mol$\%$ of ceria, achieving a volume percentage of tetragonal zirconia exceeding $65\%$ when aged at temperatures below $1000\,^\circ$C with a $2$-hour dwell time. However, they noted a decrease at higher aging temperatures, resulting in an almost entirely monoclinic phase at $1500\,^\circ$C, even with ceria amounts up to $7\,$mol$\%$. In our study, sintering via spark plasma sintering allowed us to use a high heating rate and apply uniaxial pressure, improving densification and consequently reducing grain growth while enhancing the retention of tetragonal zirconia compared to conventional sintering~\cite{anselmi2006fast,rajeswari2010comparative,cruz2012nanostructured,chen2003preparation,guillon2014field}. This allowed us to surpass a $40\,$wt.$\%$ of retained tetragonal zirconia in the mortar at a $T_{\mathrm{dwell}}$ of $1750\,^\circ$C. The use of high-pressure spark plasma sintering could further increase the quantity of retained tetragonal zirconia~\cite{maglia2010synthesis,zhang2011highly}.

\begin{figure}
	\centering
	\includegraphics[width=0.47\textwidth]{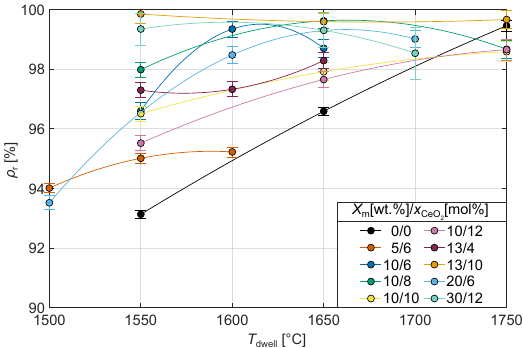}
	\caption{Final relative density of tested compositions. Variation of the final relative density ($\rho_{\mathrm{r}}$) as a function of the dwell temperature ($T_{\mathrm{dwell}}$) for different tested combinations of mortar mass ratio ($X_{\mathrm{m}}$) and ceria mole fraction ($x_{\mathrm{CeO}_2}$). Error bars represent one standard deviation.}
	\label{fig:density}
\end{figure}

Figure~\ref{fig:density} illustrates the final relative density ($\rho_{\mathrm{r}}$) values as a function of $T_{\mathrm{dwell}}$, calculated on the basis of the sample mass densities from Table~S2, phase weight fractions from Table~S1 in the Supporting Information, and densities of $\alpha$-Al$_2$O$_3$, t-ZrO$_2$, and m-ZrO$_2$ set to $3.965$, $6.10$, and $5.83\,\mathrm{g\,cm^{-3}}$, respectively~\cite{perry2016handbook,murti2022structural}. In Figure~\ref{fig:density}, the error bars were estimated using equation~\eqref{eq:sample_mass_density} and equation~\eqref{eq:theoretical_density} (Experimental Section), based on the scale resolution and the standard deviation of the phase amounts obtained from the refinement of the XRD spectra.
\begin{figure}
	\centering
	\includegraphics[width=\textwidth]{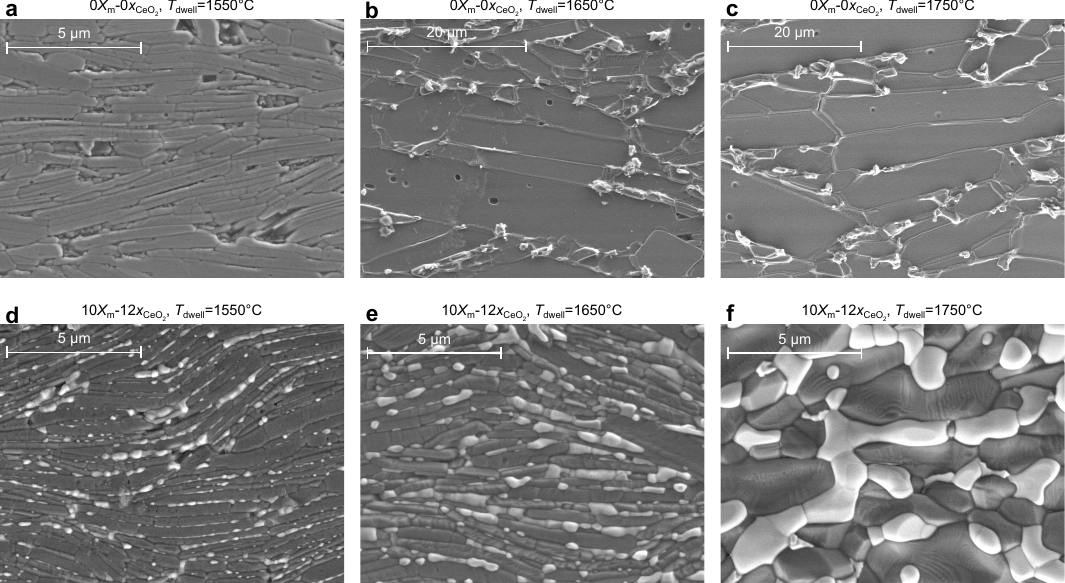}
	\caption{Microstructural analysis. SEM micrographs revealing the microstructures of samples with compositions $0X_{\mathrm{m}}$-$0x_{\mathrm{CeO}_2}$ (a-c) and $10X_{\mathrm{m}}$-$12x_{\mathrm{CeO}_2}$ (d-f), sintered at varying dwell temperature ($T_{\mathrm{dwell}}$) of $1550$ (a, d), $1650$ (b, e), and $1750\,^\circ$C (c, f). The images highlight how the introduction of mortar constrains the growth of alumina platelets. As $T_{\mathrm{dwell}}$ increases, the microstructure gradually loses its distinctive "brick and mortar" pattern. Thermal etching was performed at a temperature $150\,^\circ$C below the corresponding $T_{\mathrm{dwell}}$. SEM images were captured under low vacuum conditions with a chamber pressure of $0.6\,$bar, working distance of $5\,$mm, beam voltage of $5\,$kV and a spot size of $2.5$.}
	\label{fig:microstructures}
\end{figure}
\begin{figure}
	\centering
	\includegraphics[width=\textwidth]{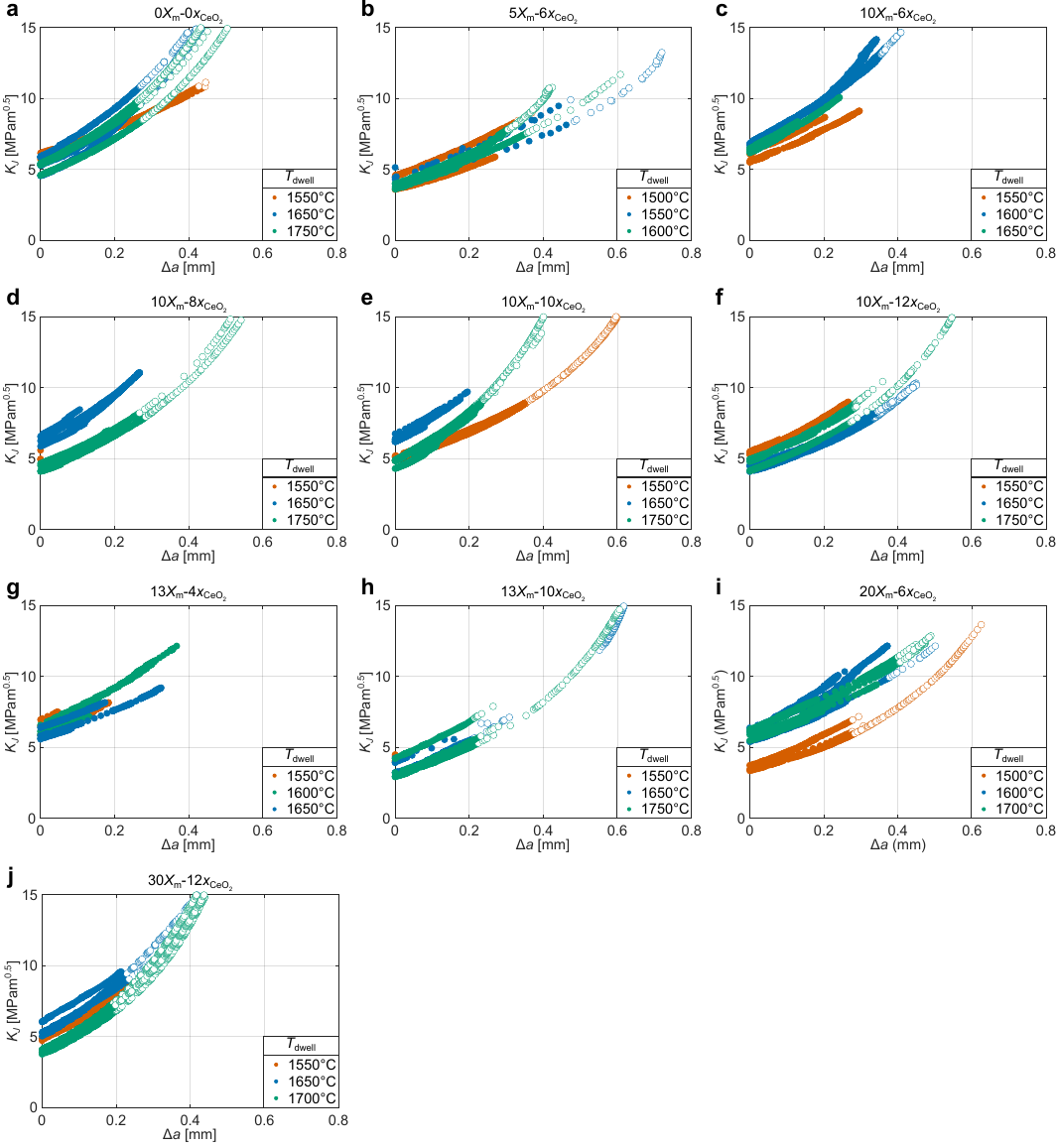}
	\caption{R-curves of tested compositions. R-curves for different tested combinations of mortar mass ratio ($X_{\mathrm{m}}$), ceria mole fraction ($x_{\mathrm{CeO}_2}$), and dwell temperature ($T_{\mathrm{dwell}}$). Unfilled markers indicate data points exceeding the crack extension limit of $0.25(W-a_0)$~\cite{iso12135}.}
	\label{fig:r_curves}
\end{figure}

The trend of $\rho_{\mathrm{r}}$ with varying $T_{\mathrm{dwell}}$ showed distinct variations among the tested compositions. In general, $\rho_{\mathrm{r}}$ initially increased with the rise in $T_{\mathrm{dwell}}$, except for specimens with compositions $13X_{\mathrm{m}}$-$10x_{\mathrm{CeO}_2}$ and $30X_{\mathrm{m}}$-$12x_{\mathrm{CeO}_2}$, where $\rho_{\mathrm{r}}$ values surpass $99\%$ at a $T_{\mathrm{dwell}}$ of $1550\,^\circ$C. Subsequent increases in $T_{\mathrm{dwell}}$ lead to a slight reduction in $\rho_{\mathrm{r}}$. Following the initial increase in $\rho_{\mathrm{r}}$, specimens with compositions $10X_{\mathrm{m}}$-$6x_{\mathrm{CeO}_2}$ and $10X_{\mathrm{m}}$-$8x_{\mathrm{CeO}_2}$ exhibited a subsequent decrease at higher $T_{\mathrm{dwell}}$ values. The reduction in $\rho_{\mathrm{r}}$ at high $T_{\mathrm{dwell}}$ may be attributed to the tetragonal to monoclinic transformation in the zirconia phase which, leading to an expansion of the grain and the formation of microcracks, results in a decrease in the final relative density~\cite{matsui2023ultrahigh}.	Notably, the onset of decreasing in $\rho_{\mathrm{r}}$ occurred at a lower $T_{\mathrm{dwell}}$ for the specimen with composition $10X_{\mathrm{m}}$-$6x_{\mathrm{CeO}_2}$ compared to the $10X_{\mathrm{m}}$-$8x_{\mathrm{CeO}_2}$. The lower retentive effect of tetragonal zirconia due to the low ceria content may lead to a rapid transformation into monoclinic with increasing $T_{\mathrm{dwell}}$.

The addition of mortar significantly influenced $\rho_{\mathrm{r}}$, with a general trend of increased mortar amounts promoting densification. Notably, specimens with composition $0X_{\mathrm{m}}$-$0x_{\mathrm{CeO}_2}$ exhibited lower $\rho_{\mathrm{r}}$ values than those with a non-zero mortar percentage for $T_{\mathrm{dwell}}$ less than $1750\text{\textdegree}$C.  
A comparison of the effect of mortar addition on $\rho_{\mathrm{r}}$ at the same ceria concentration can be made among the specimens with compositions $5X_{\mathrm{m}}$-$6x_{\mathrm{CeO}_2}$, $10X_{\mathrm{m}}$-$6x_{\mathrm{CeO}_2}$, and $20X_{\mathrm{m}}$-$6x_{\mathrm{CeO}_2}$.
The specimen with composition $10X_{\mathrm{m}}$-$6x_{\mathrm{CeO}_2}$ achieved the highest $\rho_{\mathrm{r}}$ value compared to those with $5\,$wt.$\%$ and $20\,$wt.$\%$ of mortar. Specifically, the specimen with composition $5X_{\mathrm{m}}$-$6x_{\mathrm{CeO}_2}$ displayed low $\rho_{\mathrm{r}}$ values, likely due to non-uniform mortar thickness, as highlighted in the Section 'Pre-optimization material processing', through micrograph analysis (Figure~\ref{fig:pre_optimization_analysis}~(h)). Meanwhile, the specimen with composition $20X_{\mathrm{m}}$-$6x_{\mathrm{CeO}_2}$ likely exhibited this behavior due to the considerable amount of tetragonal zirconia transforming into monoclinic during sintering~\cite{matsui2023ultrahigh}. It was thus important to find a balance for the mortar amount and composition in order to maximize densification without triggering the martensitic transformation.

A comparison, instead, of the effect of ceria addition on $\rho_{\mathrm{r}}$ with the same mortar amount can be made among the specimens with compositions $10X_{\mathrm{m}}$-$6x_{\mathrm{CeO}_2}$, $10X_{\mathrm{m}}$-$8x_{\mathrm{CeO}_2}$, $10X_{\mathrm{m}}$-$10x_{\mathrm{CeO}_2}$, and $10X_{\mathrm{m}}$-$12x_{\mathrm{CeO}_2}$. As ceria concentration increased up to $8\,$mol$\%$, an increase in the maximum $\rho_{\mathrm{r}}$ measured in the tested $T_{\mathrm{dwell}}$ was observed. A further increase in ceria concentration resulted in a decrease in the maximum $\rho_{\mathrm{r}}$ achieved in the tested $T_{\mathrm{dwell}}$.
In the literature, there are different explanations for the effects of ceria addition on the final relative density. The increase in $\rho_{\mathrm{r}}$ with increasing ceria content, as observed in zirconia-ceria powder~\cite{quinelato2000effect} and zirconia-toughened alumina~\cite{sarker2022impacts, rejab2013effects}, is attributed to the higher mass density of ceria and an increase in the retained tetragonal zirconia phase within the final microstructure, which is denser than the monoclinic counterpart.	The potential decline in $\rho_{\mathrm{r}}$ at higher ceria concentrations was associated with the formation of elongated grains of CeAl$_{11}$O$_{18}$~\cite{sarker2022impacts}, a decrease in grain boundary diffusion~\cite{mitra2002effect}, or the formation of agglomerates, as reported in CeO$_2$-Y-TZP~\cite{ragurajan2014effect}. In the current study, the formation of elongated grains of CeAl$_{11}$O$_{18}$ is to be excluded, since XRD spectra and micrograph analyses revealed no evidence of elongated grains from other phases, therefore pointing towards reduced grain boundary diffusion or the presence of localized agglomerates.
\begin{figure}
	\centering
	\includegraphics[width=\textwidth]{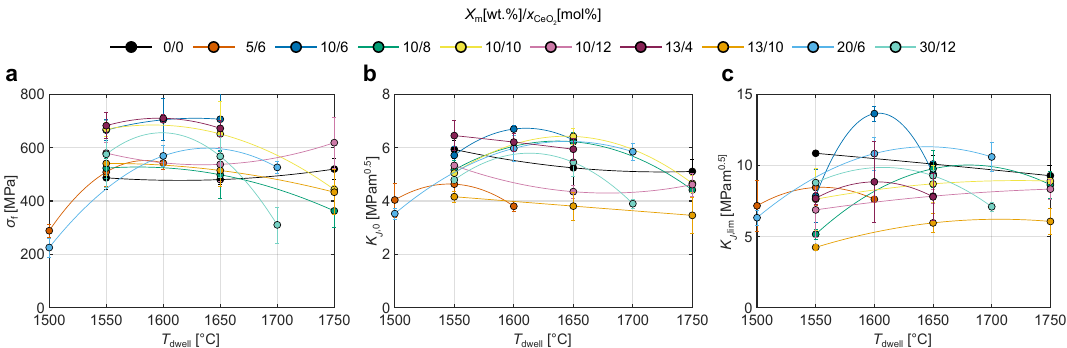}
	\caption{Mechanical properties of tested compositions. Variation in flexural strength ($\sigma_{\mathrm{f}}$) (a), stress intensity factor at crack initiation ($K_{J,\mathrm{0}}$) (b), and limit ($K_{J,\mathrm{lim}}$) (c) as a function of the dwell temperature ($T_{\mathrm{dwell}}$) for different tested combinations of mortar mass ratio ($X_{\mathrm{m}}$) and ceria mole fraction ($x_{\mathrm{CeO}_2}$). Error bars represent one standard deviation.}
	\label{fig:properties}
\end{figure}		

Figure~\ref{fig:microstructures} presents SEM micrographs illustrating the microstructures of samples with compositions $0X_{\mathrm{m}}$-$0x_{\mathrm{CeO}_2}$ (upper figures) and $10X_{\mathrm{m}}$-$12x_{\mathrm{CeO}_2}$ (lower figures), where the lighter phase is the zirconia mortar. The samples were sintered at $T_{\mathrm{dwell}}$ temperatures of $1550$, $1650$, and $1750\,^\circ$C, progressing from left to right. A comparison of Figure~\ref{fig:microstructures}~(a-c), where different bar scales were used, reveals a significant increase in grain size with the growth of $T_{\mathrm{dwell}}$.
In Figure~\ref{fig:microstructures}~(d-f), the samples were sintered at the same $T_{\mathrm{dwell}}$ temperatures as those above, emphasizing the mortar's role as a pinning phase that restricts the growth of alumina platelets~\cite{okada1992role}. However, as $T_{\mathrm{dwell}}$ increased, the microstructure underwent a transformation, losing its distinctive "brick and mortar" characteristic. At the highest $T_{\mathrm{dwell}}$, it exhibited agglomerates of zirconia grains dispersed in an alumina matrix.

\section{Influence of Process Parameters on Mechanical Properties}	
We calculated the R-curves for the tested compositions using equation~\eqref{eq:stress_intensity_factor} to equation~\eqref{eq:equivalent_stress_intensity_factor} (Experimental Section). In equation~\eqref{eq:j_integral}, we determined the values of $E$ and $\nu$ using the rule of mixture~\cite{ashby2011material}, assuming the mortar and bricks underwent the same strain. The elastic modulus and Poisson's ratio of alumina perpendicular to the c-axis were taken as $431\,$MPa, and $0.294$~\cite{yousef2005microcrack}, and those of polycrystalline ceria-stabilized zirconia were considered $195\,$MPa, and $0.32$~\cite{pabst2004effective}. In Figure~\ref{fig:r_curves}~(a-j), the R-curves are illustrated, derived from three-point bending tests on notched specimens for the same composition at different $T_{\mathrm{dwell}}$, with three repetitions for each $T_{\mathrm{dwell}}$. Filled markers indicate points within the crack extension limit~\cite{iso12135}. Some compositions, such as $10X_{\mathrm{m}}$-$8x_{\mathrm{CeO}_2}$ and $13X_{\mathrm{m}}$-$10x_{\mathrm{CeO}_2}$, both sintered at $1550\,^\circ$C, did not exhibit typical R-curve behavior. In these instances, there was no stable crack propagation phase, and the R-curve degenerated to a point on the ordinate axis, possibly due to a high flaw concentration resulting from low final relative density.
The stress intensity factors for both crack initiation ($K_{J,\mathrm{0}}$) and limit ($K_{J,\mathrm{lim}}$), along with the flexural strength ($\sigma_{\mathrm{f}}$), obtained from three-point bending tests on smooth specimens are summarized in Table~S2 in the Supporting Information. The standard deviation reported in the table was calculated based on three tests conducted on different samples.

Figure~\ref{fig:properties}~(a) shows the variation of $\sigma_{\mathrm{f}}$ as a function of $T_{\mathrm{dwell}}$ for the tested compositions. The $\sigma_{\mathrm{f}}$ values of the sample with composition $0X_{\mathrm{m}}$-$0x_{\mathrm{CeO}_2}$ remained approximately constant as the $T_{\mathrm{dwell}}$ varied. Although the final relative density of the composition $0X_{\mathrm{m}}$-$0x_{\mathrm{CeO}_2}$ showed a significant increase as the $T_{\mathrm{dwell}}$ increased (Figure~\ref{fig:density}), the reduction of pore defects associated with the increase in final relative density did not lead to an improvement in mechanical properties due to the increase in grain size (Figure~\ref{fig:microstructures}).
Adding mortar increased $\sigma_{\mathrm{f}}$ in the $T_{\mathrm{dwell}}$ range between $1550$ and $1650\,^\circ$C. At $1750\,^\circ$C, all compositions with mortar addition exhibited a decrease in $\sigma_{\mathrm{f}}$, except for the composition $10X_{\mathrm{m}}$-$12x_{\mathrm{CeO}_2}$, which showed almost constant $\sigma_{\mathrm{f}}$ as $T_{\mathrm{dwell}}$ varied. Several factors could contribute to strength reduction at elevated dwell temperatures, including grain growth or the formation of microcracks due to tetragonal to monoclinic phase transformation during cooling. The highest $\sigma_{\mathrm{f}}$ values were measured for sample $13X_{\mathrm{m}}$-$4x_{\mathrm{CeO}_2}$, which had the lowest ceria content, at temperatures of $1600\,^\circ$C.

Stress intensity factors at both crack initiation and limit, as functions of $T_{\mathrm{dwell}}$, are shown in Figure~\ref{fig:properties}~(b), and (c).
The tested compositions generally exhibited an initial increase in both $K_{J,\mathrm{0}}$ and $K_{J,\mathrm{lim}}$ as $T_{\mathrm{dwell}}$ increased, followed by reaching a maximum point and subsequently decreasing with further increases in $T_{\mathrm{dwell}}$. The initial rise in $K_{J,\mathrm{0}}$ and $K_{J,\mathrm{lim}}$ with increasing $T_{\mathrm{dwell}}$ was expected due to the reduction in material porosity. With a further increase in $T_{\mathrm{dwell}}$, excessive grain growth triggered a tetragonal to monoclinic phase transformation in the zirconia phase during sintering. This transformation induced microcracks and reduced transformation toughening due to a lower amount of tetragonal zirconia in the final composition. These phenomena were highlighted in the Section 'Influence of process parameters on phase composition, microstructure and density', where the effects of increasing $T_{\mathrm{dwell}}$ on the final relative density (Figure~\ref{fig:density}) and on the amount of tetragonal zirconia in the resulting composition (Figure~\ref{fig:xrd_results}~(c)) were observed.	
The decrease in stress intensity factors was not observed in specimens with compositions $0X_{\mathrm{m}}$-$0x_{\mathrm{CeO}_2}$, $10X_{\mathrm{m}}$-$12x_{\mathrm{CeO}_2}$, and $13X_{\mathrm{m}}$-$10x_{\mathrm{CeO}_2}$ for both $K_{J,\mathrm{0}}$ and $K_{J,\mathrm{lim}}$, and in $10X_{\mathrm{m}}$-$10x_{\mathrm{CeO}_2}$ for $K_{J,\mathrm{lim}}$. Specifically, the specimens with composition $0X_{\mathrm{m}}$-$0x_{\mathrm{CeO}_2}$ lack transformation toughening since they did not have a zirconia phase. The other compositions were characterized by a stable phase of tetragonal zirconia. In fact, they possessed the highest quantity of tetragonal zirconia among the tested compositions, as highlighted in Figure~\ref{fig:xrd_results}~(c), and the tetragonal to monoclinic phase transformation induced by sintering was limited even at high values of $T_{\mathrm{dwell}}$. Consequently, $\rho_{\mathrm{r}}$ did not exhibit reductions (Figure~\ref{fig:density}), and $K_{J,\mathrm{0}}$ and $K_{J,\mathrm{lim}}$ did not show sharp drops with increasing $T_{\mathrm{dwell}}$.	
It is noteworthy that the high stability of the tetragonal phase increased the stress necessary for the stress-induced tetragonal to monoclinic phase transformation~\cite{sharma2002synthesis,hannink2000transformation}, resulting in limited values of stress intensity factors.

At a temperature of $1600\,^\circ$C, the specimen with composition $10X_{\mathrm{m}}$-$6x_{\mathrm{CeO}_2}$, demonstrated the highest $K_{J,\mathrm{lim}}$ value. The $K_{J,\mathrm{lim}}$ value achieved was extremely high ($13.6\,\mathrm{MPam^{0.5}}$). It corresponded to a $400\%$ increase in the micro-scale fracture toughness of alumina grains in textured alumina ceramics~\cite{schlacher2023micro}. Although, as shown in Figure~\ref{fig:data_comparison}, there are nacre-like composites with higher toughness, they possess significantly lower strength. We can attribute a contribution to the toughness to the stress-induced tetragonal to monoclinic phase transformation. Indeed, the specimen with the composition $10X_{\mathrm{m}}$-$6x_{\mathrm{CeO}_2}$ exhibited a large crystal size at $1600\,^\circ$C, as highlighted in Figure~\ref{fig:xrd_results}~(a). An increase in $T_{\mathrm{dwell}}$ induced a rapid and spontaneous transformation of the tetragonal phase, as evidenced by the significant reduction in tetragonal zirconia content with the rise of $T_{\mathrm{dwell}}$ in Figure~\ref{fig:xrd_results}~(c). This suggests that, at this composition, the crystal size approached the critical value for a spontaneous to monoclinic phase transformation. A crystal size close to the critical value requires lower strain energy density to induce phase transformation upon external stress~\cite{garvie1985part1,garvie1985part2}, contributing to improved mechanical properties~\cite{bravo2002fracture}. Nevertheless, we cannot exclude the presence of other toughening phenomena, as tetragonal zirconia constitutes a part of the final composition.
It is also worth noting that monoclinic zirconia, as observed in the literature, can contribute to the increase in toughness in zirconia-toughened alumina through the nucleation of microcracks that occur in the stress field of propagating cracks~\cite{ruehle1986transformation}.
\begin{figure}
	\centering
	\includegraphics[width=\textwidth]{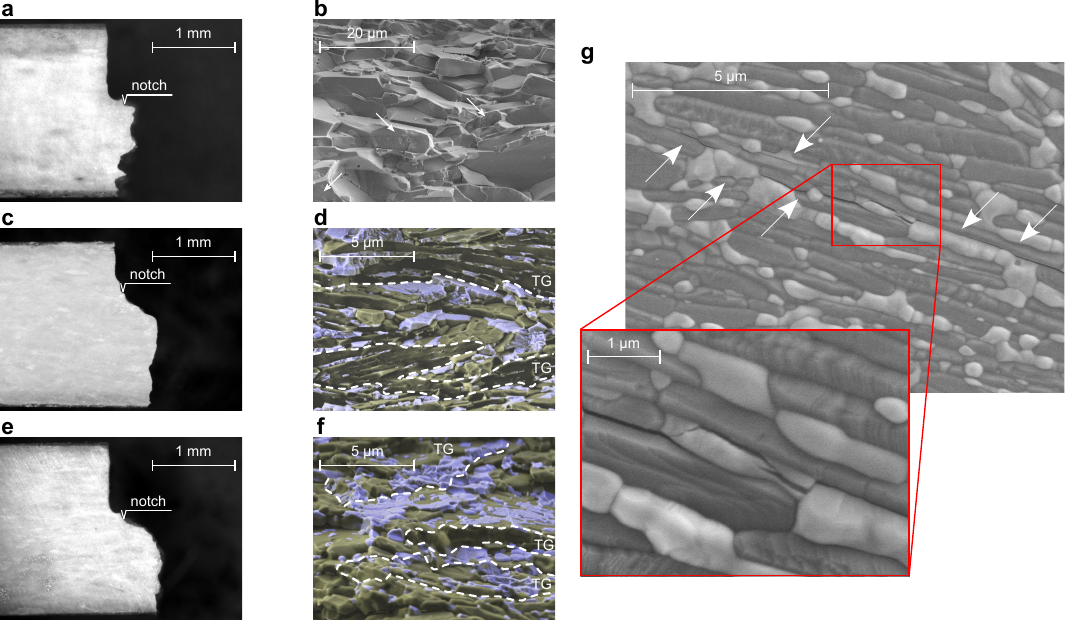}
	\caption{Fractographic Analysis. Optical micrographs capturing the crack of the notched specimens (a) $0X_{\mathrm{m}}$-$0x_{\mathrm{CeO}_2}$, (c) $10X_{\mathrm{m}}$-$10x_{\mathrm{CeO}_2}$ sintered at a dwell temperature of $1650\,^\circ$C, and (e) $10X_{\mathrm{m}}$-$6x_{\mathrm{CeO}_2}$, sintered at a dwell temperature of $1600\,^\circ$C, along with the corresponding SEM fractographs (b, d, f). In (b), arrows indicate regions of platelet fracture, while in (d, f), large areas predominantly exhibiting transgranular fracture, where the crack passes through both platelets and mortar, are labeled as "TG". (g) displays a secondary crack starting from the main crack of the notched specimen after testing the material with composition $20X_{\mathrm{m}}$-$6x_{\mathrm{CeO}_2}$ sintered at $1700\,^\circ$C, with arrows indicating the path of the crack focusing on the point where it passes through the mortar phase. SEM images are obtained under low vacuum conditions with a chamber pressure of $0.6\,$bar, working distance of $5\,$mm, beam voltage of $5\,$kV, and a spot size of $2.5$. The fractographs (d, f) are composite images combining results from a large field and back-scattered electron detector, accentuating the zirconia phase in light blue.}
	\label{fig:fractographs}
\end{figure}

Figure~\ref{fig:fractographs} displays the crack paths after mechanical testing in two notched specimens with compositions (a, b) $0X_{\mathrm{m}}$-$0x_{\mathrm{CeO}_2}$ and (c, d) $10X_{\mathrm{m}}$-$10x_{\mathrm{CeO}_2}$, both sintered at a $T_{\mathrm{dwell}}$ of $1650\,^\circ$C, along with those of the specimen with composition $10X_{\mathrm{m}}$-$6x_{\mathrm{CeO}_2}$, sintered at a $T_{\mathrm{dwell}}$ of $1600\,^\circ$C, which exhibited the highest toughness (e, f).
The crack propagation paths in these samples represent the behavior observed in samples with and without mortar. The sample with composition $0X_{\mathrm{m}}$-$0x_{\mathrm{CeO}_2}$ showed repeated crack deflections until it passed through the entire sample. In contrast, the samples with composition $10X_{\mathrm{m}}$-$10x_{\mathrm{CeO}_2}$, and $10X_{\mathrm{m}}$-$6x_{\mathrm{CeO}_2}$ initially showed marked crack deflection, then aligning with the loading direction and exhibiting smaller deflections.
The fractography in Figure~\ref{fig:fractographs}~(b) revealed a mixed-type crack. It presents regions of intergranular fracture where the crack was deflected, crossing grain boundaries of the large platelets grown during sintering, and localized regions of transgranular fracture highlighted by arrows where the crack passed through the platelets.
Figure~\ref{fig:fractographs}~(d, f), acquired in a region far from the notch, demonstrated that the crack was characterized by regions with different features. In the regions labeled as "TG," the crack predominantly displayed transgranular characteristics in both zirconia and alumina. It traversed the platelets parallel to the crystallographic c-axis, leading to zirconia fracture with a rough appearance. In other regions, the crack deviated as a result of the "brick and mortar" microstructure, showing a mix of transgranular and intergranular characteristics and regions of failure at the alumina-zirconia interface.
Figure~\ref{fig:fractographs}~(g) depicts secondary cracks branching out from the main crack of the notched specimen after testing the material with composition $20X_{\mathrm{m}}$-$6x_{\mathrm{CeO}_2}$ sintered at $1700\,^\circ$C. Arrows indicate the path of the crack as it traversed through the material at the brick-mortar interface, with regions where it intersected with the mortar phase. Multiple cracks were present, aiding in the dissipation of crack energy and contributing to the material's toughness.
Crack deflection in samples containing mortar dominates in the initial phase of crack propagation. However, stable crack propagation throughout the entire mechanical test was reasonably attributed to transformation toughening phenomena during the entire crack propagation.

Crack deflection was observed in the material with composition $10X_{\mathrm{m}}$-$6x_{\mathrm{CeO}_2}$ sintered at a $T_{\mathrm{dwell}}$ of $1600\,^\circ$C (Figure~\ref{fig:fractographs}~(e)), which recorded the maximum value of the objective function. Its high toughness was attributed to a combination of two main toughening mechanisms: transformation toughening and crack deflection, along with a flexural strength comparable to the maximum measured among the compositions tested.

\section{Conclusions}
\label{section:conclusions}
We engineered an ultra-strong "double-tough" ceramic consisting of nacre-like alumina with ceria-stabilized zirconia as a mortar phase. Utilizing a sol-gel technique, we successfully produced nanograined tetragonal zirconia uniformly distributed around the platelets. After spark plasma sintering, we achieved bulk materials with a brick-and-mortar microstructure, with the platelets oriented in a preferred direction. To expedite the experimental campaign aimed at finding the optimal process parameters (ceria concentration,  mortar amount, and dwell temperature), we devised a batch Bayesian optimization framework accounting for the cost benefits of making materials with the same composition, where the goal was to maximize a balance between strength and toughness.	

Leveraging both transformation toughening at the nanoscale and crack deflection at the microscale, the resulting materials achieved an exceptional combination of fracture toughness and strength, with record-high flexural strength for nacre-like aluminas ($>700\,$MPa) and very high fracture toughness ($>13\,\mathrm{MPa\,m^{0.5}}$). Notably, the only reported nacre-like alumina surpassing this fracture toughness values has strength levels well below $500\,$MPa. Our BO-based approach allowed us to identify the optimal process parameters to achieve this combination of properties. This can be further fine-tuned by assigning different weights to strength and toughness in the BO objective function.
Additional insights are:
\begin{itemize}
	\item The zirconia mortar contributes to the densification of the material, particularly when the dwell temperature remains low enough to prevent excessive tetragonal to monoclinic transformation during cooling. Additionally, it acts as a pinning phase, limiting the growth of alumina grains and resulting in increased flexural strength at dwell temperatures between $1550$ and $1650\,^\circ$C.
	\item The increase in dwell temperature enhances the final relative density of the material with and without mortar until it reaches a maximum value. Beyond this point, there is a reduction in final relative density in almost all tested compositions. At excessively high dwell temperatures, the all-ceramic composite tends to lose its characteristic "brick-and-mortar" structure, with its microstructure appearing as dispersed zirconia grains in an alumina matrix. Moreover, the size of the zirconia crystallites increases, promoting the tetragonal to monoclinic transformation during cooling and reducing the retained tetragonal zirconia, thereby decreasing transformation toughening.
	\item The tetragonal to monoclinic transformation during cooling is associated with a general reduction in mechanical properties.
	\item The use of spark plasma sintering facilitated the attainment of high levels of tetragonal zirconia in the final composition, surpassing $40\,$wt.$\%$ of retained tetragonal zirconia in the mortar at temperatures of $1750\,^\circ$C.
	\item Ceria serves as a stabilizer for the tetragonal phase, increasing the quantity of retained tetragonal zirconia with increasing dwell temperature. At high ceria contents, the unfavorable transformation of tetragonal zirconia to monoclinic results in limited toughness due to a lack of contribution from transformation toughening.
\end{itemize}

The proposed process optimization framework not only expedites the material design but also serves as a promising pathway for the development of materials with tailored and improved mechanical properties. 
We foresee that exploring additional process parameters - such as the size and aspect ratio of the alumina platelets, different sintering temperature and pressure profiles, or alternative stabilizing oxides for the zirconia phase - can further optimize the material's mechanical performance. The applicability of the proposed methodology also extends well beyond the concept of double-tough ceramics, serving as valuable optimization framework for a variety of cumbersome, time and energy-intensive experimental processes.


\section{Experimental Section}
\label{section:material-and-methods}
This section outlines the process procedure designed to create nacre-like alumina with a compliant phase of ceria-stabilized zirconia. We elaborate on the methodologies used for both microstructural and mechanical characterization, along with the procedural aspects employed to optimize the resulting mechanical properties.

\subsection{Materials}
\label{subsection:materials}
To fabricate the "brick" phase, we used commercial Al$_2$O$_3$ platelets (Serath YFA10030BJ) supplied by Kinsei Matek Co., Ltd. (Osaka, Japan), featuring an average particle size of $10\,\mathrm{\mu m}$ with fine particles removed, and an aspect ratio ranging from $25$ to $30$.
For the creation of the "mortar" phase, we employed the following materials: a $70\,\mathrm{wt.\%}$ solution of zirconium(IV) propoxide, Zr(OC$_3$H$_7$)$_4$, in $1$-propanol (with a density of $1.044\,\mathrm{g\,cm^{-3}}$ at $25\,^\circ$C), cerium(III) nitrate hexahydrate, Ce(NO$_3$)$_3\!\cdot\!6$H$_2$O, of $99\%$ trace metals basis, and nitric acid, HNO$_3$, with a concentration exceeding $65\%$. These materials were sourced from Merck KGaA (Darmstadt, Germany).

\subsection{Processing of the bulk double-tough ceramics}
\label{subsection:material_processing}
We made the material by coating Al$_2$O$_3$ platelets with ceria-stabilized zirconia. Subsequently, we spark plasma sintered the resulting powder to obtain fully dense bulk materials with the desired microstructure. For the coating procedure, we used sol-gel methods, following a methodology based on the one outlined in~\cite{shen2010synthesis,rossignol1999preparation}. 

We prepared a water solution of ceria-stabilized zirconia precursors for the coating step. For each liter of water, we added $71.57\,\mathrm{ml}$ of Zr(OC$_3$H$_7$)$_4$ reagent (denoted as $V_{\mathrm{{Zr(reag.)}}}$), $30\,\mathrm{ml}$ of HNO$_3$ reagent, and a mass, $m_{\mathrm{Ce(reag.)}}$, of Ce(NO$_3$)$_3\!\cdot\!6$H$_2$O reagent. We achieve different ceria mole fractions, $x_{\mathrm{CeO}_2}$, in the coating by adjusting $m_{\mathrm{Ce(reag.)}}$. We determined $m_{\mathrm{Ce(reag.)}}$ through stoichiometric calculations using
\begin{equation}
	m_{\mathrm{Ce(reag.)}}=n_{\mathrm{Ce}}\frac{100}{p_{\mathrm{Ce(reag.)}}}M_{\mathrm{Ce(NO_3)_3\cdot6H_2O}}, 
	\label{eq:mass_Ce_reag}
\end{equation}
with
\begin{equation}
	n_{\mathrm{Ce}}=\frac{x_{\mathrm{CeO}_2}}{1-x_{\mathrm{CeO}_2}}n_{\mathrm{Zr}},
	\label{eq:mol_Ce}
\end{equation}
\begin{equation}
	n_{\mathrm{Zr}}=V_{\mathrm{{Zr(reag.)}}}\frac{{c_{\mathrm{Zr(reag.)}}}}{100}\frac{\rho_{\mathrm{{Zr(reag.)}}}}{M_{\mathrm{Zr(OC_3H_7)_4}}},
	\label{eq:mol_Zr}
\end{equation}
where $n$ represents the amount of the element in solution (in mol), $M$ the molar mass, $p$ the reagent purity, $c$ the reagent weight concentration, and $\rho$ the mass density; each symbol refers to its subscripts. Upon the introduction of the Zr(OC$_3$H$_7$)$_4$ reagent, the mixture underwent hydrolysis, leading to the formation of an initially opaque mixture. We subjected this mixture to stirring at room temperature on a magnetic stirrer at $500\,$rpm for a duration of $24\,$hours, resulting in the solution becoming transparent.

After preparing the coating solution, we coat the Al$_2$O$_3$ platelets. Initially, we added $10\,$g of Al$_2$O$_3$ platelets to $180\,$ml of distilled water, followed by a $15$-minute sonication process. Subsequently, we introduced a quantity of the coating solution to achieve the final "mortar" mass ratio, $X_{\mathrm{m}}$ defined as
\begin{equation}
	X_{\mathrm{m}}=\frac{m_{\mathrm{m}}}{m_{\mathrm{Al_2O_3}}+m_{\mathrm{m}}},
\end{equation}
with $m_{\mathrm{m}}$ is the "mortar" mass, and $m_{\mathrm{Al_2O_3}}$ the alumina platelets mass ("bricks"). We calculated the coating solution volume to add under the assumption that all the added zirconium and cerium would undergo complete oxidation during calcination. Given the molarity of zirconium in the prepared coating solution (approximately $0.16\,\mathrm{mol\,L^{-1}}$), we selected a volume of coating solution containing the required moles of Zr, $n_{\mathrm{Zr}}$, determined by
\begin{equation}
	n_{\mathrm{Zr}}=(1-x_{\mathrm{CeO}_2})\frac{\frac{X_{\mathrm{m}}}{1-X_{\mathrm{m}}}m_{\mathrm{Al_2O_3}}}{(1-x_{\mathrm{CeO}_2})M_{\mathrm{ZrO_2}}+x_{\mathrm{CeO}_2}M_{\mathrm{CeO_2}}}.
\end{equation}
After adding the coating solution, we mixed it on a hot plate at a temperature of $70\,^\circ$C for $5$ hours, followed by an additional $19$ hours at room temperature. Subsequently, we dried the solution and deagglomerated the coated platelets using a mortar and pestle.

We subjected the obtained powder to calcination in a muffle furnace at $600\,^\circ$C for $4$ hours in an air environment. This process served the dual purpose of removing volatile substances and oxidizing the elements in the "mortar" phase.

\begin{figure}
	\centering
	\includegraphics[width=\textwidth]{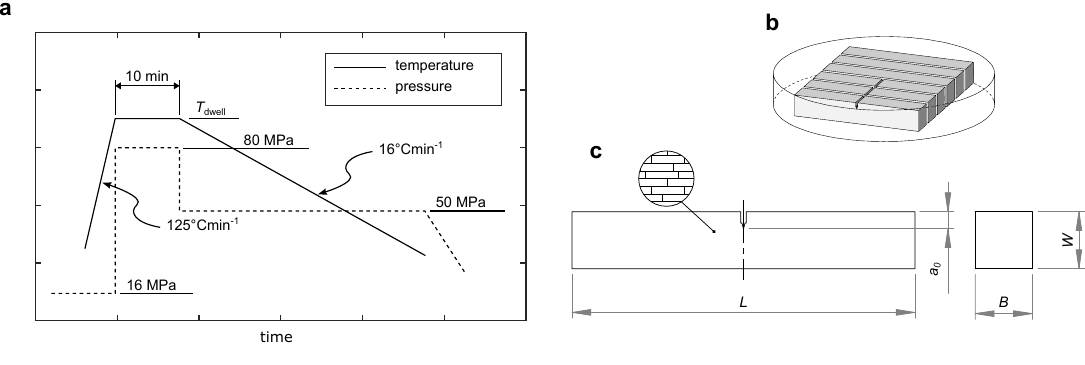}
	\caption{Spark plasma sintering conditions and specimen geometry for mechanical testing. Temperature and pressure profiles employed during spark plasma sintering (a), specimen extraction configuration for three notched and three smooth specimens from the sintered samples (b), and detailing the single edge V-notch specimen geometry, including microstructure orientation, key geometric parameters, and their corresponding nomenclature (c).}
	\label{fig:methods}
\end{figure}

For the subsequent sintering process, we used $3.5\,$g samples of the calcined powder and employed the spark plasma sintering technique~\cite{guillon2014field,giuntini2013localized}. The samples were placed within $20\,$mm diameter graphite dies. Additionally, a $0.3\,$mm graphite sheet was placed onto the side surface of the die, and two circular sheets were inserted between the sample and the die bases. The sintering operation was conducted under vacuum conditions. The applied sintering cycle, as illustrated in Figure~\ref{fig:methods}~(a), involved a constant heating rate of $125\,^\circ\mathrm{C\,min^{-1}}$ until reaching the maximum temperature, denoted as $T_{\mathrm{dwell}}$.
Subsequently, we maintained this constant temperature for $10\,$min, followed by a controlled cooling phase with a rate of $16\,^\circ\mathrm{C\,min^{-1}}$. Throughout the ramp-up to the maximum temperature, we maintained a pressure of $16\,$MPa. During the  temperature hold period, we increased the pressure to $80\,$MPa. To prevent the formation of cracks in the material and safeguard the equipment, we then reduced the pressure to $50\,$MPa during the cooling phase. 
We determined the final mass density of the sintered sample, $\rho_{\mathrm{s}}$, using Archimedes method. We measured the sample's mass both in air, $m_{\mathrm{a}}$, and when immersed in water, $m_{\mathrm{w}}$. In our calculations, we neglected the buoyancy of the sample in air. We calculated $\rho_{\mathrm{s}}$ using~\cite{spierings2011comparison}
\begin{equation}
	\rho_{\mathrm{s}}=\frac{m_{\mathrm{a}}}{m_{\mathrm{a}}-m_{\mathrm{w}}}\rho_{\mathrm{w}},
	\label{eq:sample_mass_density}
\end{equation}
with $\rho_{\mathrm{w}}=0.998\,\mathrm{g\,cm^{-3}}$ denoting the mass density of water at a temperature of $20\,^\circ$C and a pressure of $1\,$atm~\cite{nist2008thermophysical}. We calculated the final relative density of the sample, $\rho_{\mathrm{r}}$, by dividing $\rho_{\mathrm{s}}$ by the theoretical density, $\rho_{t}$, obtained from the rule of mixture~\cite{liu2008relationship}
\begin{equation}
	\frac{1}{\rho_{t}}=\sum_i\frac{\chi_i}{\rho_i},
	\label{eq:theoretical_density}
\end{equation}
where $\chi_i$ is the weight fraction of the $i$-th phase and $\rho_i$ its mass density.

\subsection{Micrographic analysis}
We conducted micrographic analyses employing a scanning electron microscope (SEM) equipped with an energy dispersion x-ray analyzer (EDX) to assess the effectiveness of platelet coating, examine the final material microstructure, and inspect the fracture surfaces of tested specimens. We use low vacuum imaging at a chamber pressure of $0.6\,$mbar to reduce charging, with a working distance of $5\,$mm, beam voltage ranging between $5$ and $10\,$kV, and a spot size between $2.5$ and $3$. 

For sample preparation, we embedded a portion of the material in a $30\,$mm diameter capsule using the thermosetting bakelite resin with carbon filler PolyFast supplied by Struers. Subsequently, we conducted automated grinding using abrasive papers with grit sizes ranging from $320$ to $4000$~grit, gradually increasing grinding times. During this process, we used water as a cooling fluid, applied a force of $35\,$N to each specimen, and maintained a polishing head speed of $300\,$rpm, rotating in the same direction as the polishing platen. Following grinding, we polished the samples using a $1\,\mathrm{\mu m}$ diamond suspension. For this operation, we reduced the rotation speed of the platen and head by half.

Furthermore, we subjected polished samples to thermal etching. This process entailed removing them from the mounting resin and placing them in a furnace set at a temperature $150\,^\circ$C lower than $T_{\mathrm{dwell}}$, with an etching duration of $30$ minutes~\cite{chinn2002preparation}.

\subsection{X-ray diffraction analysis}
We conducted an X-ray diffraction (XRD) analysis using a Bruker D2 PHASER XE-T instrument equipped with a Co-anode X-ray source. We analyzed the material at various stages of the process, including before and after calcination, as well as after sintering. The XRD measurements covered a range of diffraction angles, $2\theta$, from $25$ to $55$ degrees. We collected data points at an increment of $0.01$ degrees, with a time step of $1$ second per step. 

We used GSAS II software~\cite{toby2013gsas} for the data analysis. We initiated a crystal structure refinement process with the crystal structure models of tetragonal zirconia~\cite{igawa2001crystal}, monoclinic zirconia~\cite{fukui2019structure}, and alpha alumina~\cite{maslen1993synchrotron} sourced from the Crystallography Open Database~\cite{grazulis2009}. We performed parameter refinement for the phase quantities, sample displacement, lattice parameters, and crystallite size. To account for preferred orientation effects in crystallites, we applied a spherical harmonics model of order $2$.

\subsection{Mechanical characterization}
We conducted flexural strength and resistance to crack propagation characterizations of the sintered material at room temperature using three-point bending tests carried out on smooth and notched specimens. Figure~\ref{fig:methods}~(b) illustrates the specimen extraction configuration from the sintered sample, with a single-edge V-notch beam employed for notched specimens. Figure~\ref{fig:methods}~(c) presents the geometry of the notch specimen, highlighting microstructure orientation, key geometric parameters, and their corresponding nomenclature. Notably, the smooth specimen shares the same geometry, microstructure orientation, and parameter nomenclature as the notched one, except for the absence of a notch.

We extract six bars from the sintered cylindrical sample measuring $2\!\times\!2\!\times\!12\,\mathrm{mm}^3$  using a diamond disc saw. We finished the extracted bars with sandpaper ranging from $320$ to $800$~grit, and chamfered the long edges to a size of $0.12\,\mathrm{mm} \pm 0.03\,\mathrm{mm}$ at a $45^{\circ}$ to prevent the formation of cracks from sharp edges during testing. Among the six specimens extracted, we notched three of them. We temporarily fixed the specimens side by side onto a support using adhesive. Initially, we created a starter notch using a diamond wire saw with a diameter of $0.26\,$mm. We filled the starter notch with diamond paste featuring a $1\,\mathrm{\mu m}$ particle size, and created a sharp V-notch using a single-edged surgical carbon-steel blade, applying a light back-and-forth motion. The overall length of the notch, $a_0$, ranged between $0.45$ and $0.55$ times the width, $W$.
Prior to conducting the mechanical tests, we subjected all specimens to cleaning in an ultrasonic bath with acetone. Subsequently, we measured the width ($W$) and thickness ($B$) of each specimen using micrometer calipers. After each mechanical test, we measured the initial overall notch length ($a_0$) on the fractured surface, employing a microscope equipped with a calibrated stage.

We applied three-point bending tests by positioning the specimen's $B\!\times\!L$ surface on two support pins with a span, $S$, equal to $5\,$mm. We positioned the V-notch at the midpoint between the support pins. We applied the load under displacement control to the opposite surface using a central loading pin, at a rate of $1\,\mathrm{\mu m\,s^{-1}}$. We continuously recorded the applied force, $F$, and displacement, $d$, during the entire test, at a data acquisition rate of $50\,$Hz, until complete fracture occurred.
We calculated the flexural strength, $\sigma_{\mathrm{f}}$, based on tests conducted on a smooth specimen, following the ISO~14704:2016 standard~\cite{iso14704}. The calculation assumes that the specimen behaves within a linear elastic range until reaching the maximum load, $F_{\mathrm{f}}$, which corresponds to the moment of fracture. We determined $\sigma_{\mathrm{f}}$ using~\cite{iso14704}
\begin{equation}
	\sigma_{\mathrm{f}}=\frac{3F_{\mathrm{f}}S}{2BW^2}.
	\label{eq:flexural_strength}
\end{equation}
We interpreted the $F\!-\!d$ curve obtained from tests on notched specimens to determine the material's crack growth resistance curve (R-curve). Employing the compliance method, we identified the initiation of crack propagation and crack length during the test. We defined the onset of propagation as the point at which the $F\!-\!d$ curve exhibits a change in compliance, and we calculated the crack length using the recursive relationship~\cite{saad2020simple}
\begin{equation}
	a_i=a_{i-1}+\frac{W-a_{i-1}}{2}\frac{C_i-C_{i-1}}{C_i},
	\label{eq:crack_length_compliance}
\end{equation}
with $a_i$ representing the crack length corresponding to the compliance $C_i=d_i/F_i$. We employed the $J$ contour integral as the fracture characterizing parameter for the material~\cite{anderson2017fracture}. To determine its value, denoted as $J$, we followed the procedure outlined in the ISO 12135:2021 standard~\cite{iso12135}
\begin{equation}
	J=\frac{K^2(1-\nu^2)}{E}+\frac{1.9A_{\mathrm{pl}}}{B(W-a)},
	\label{eq:j_integral}
\end{equation}
where $K$ represents the stress intensity factor, $\nu$ is the Poisson's ratio, $E$ is the modulus of elasticity, and $A_{\mathrm{pl}}$ is the plastic component of the total area, $A_{\mathrm{tot}}$, under the $F\!-\!d$ curve at the force, $F$ ($A_{\mathrm{pl}}=A_{\mathrm{tot}}-Fd/2$). While the standard prescribes a relationship for calculating $K$ for specimens with a standard geometry and a span-to-width ratio of $4$, our specimen's geometry deviates from this standard. Therefore, we utilized the formulation proposed by Guinea et al.~\cite{guinea1998stress} to calculate $K$ as a function of any span-to-width ratio greater than $2.5$. This approach underwent validation through finite element simulations and demonstrated favorable comparisons with the standard formulation. Let's define the crack-to-width ratio as $\alpha=a/W$ and the span-to-width ratio as $\beta=S/D$. Under these definitions
\begin{equation}
	K=\frac{3FS}{2BW^{3/2}} \frac{\alpha^{1/2} p_{\beta}(\alpha)}{(1-\alpha)^{3/2}(1+3\alpha)},
	\label{eq:stress_intensity_factor}
\end{equation}
where $p_{\beta}(\alpha)=p_{\infty}+4/\beta \left[p_4(\alpha)-p_{\infty}(\alpha)\right]$, with
\begin{equation}
	p_4(\alpha)=1.9+0.41\alpha+0.51\alpha^2-0.17\alpha^3,
	\label{eq:shape_factors_p4}
\end{equation}
\begin{equation}
	p_{\infty}(\alpha)=1.99+0.83\alpha-0.31\alpha^2+0.14\alpha^3.
	\label{eq:shape_factors_pinf}
\end{equation}
We represented the R-curve as the equivalent stress intensity factor, $K_J$, as a function of the crack extension $\Delta a = a-a_0$. We determined $K_J$ from $J$ using
\begin{equation}
	K_J=\sqrt{\frac{JE}{1-\nu^2}},
	\label{eq:equivalent_stress_intensity_factor}
\end{equation}
We computed two key parameters from the R-curve: the stress intensity factor at crack initiation, denoted as $K_{J,0}$, assumed for $\Delta a=0$, and at the crack extension limit, denoted as $K_{J,\mathrm{lim}}$, according the standard~\cite{iso12135} for $\Delta a=0.25(W-a_0)$.

\subsection{Material optimization}
\label{sec:material_optimization}
Our approach focuses on maximizing the objective function $f(\mathbf{x})$, where $\mathbf{x}=[x_{\mathrm{CeO}_2},\ X_{\mathrm{m}},\ T_{\mathrm{dwell}}]$. We defined the function $f\colon\mathcal{X}\rightarrow\mathbb{R}$, where $\mathcal{X}\subseteq\mathbb{R}^d$, as a linear combination of dimensionless mechanical properties $ K_{J,\mathrm{lim}}/\bar{K}_{J,\mathrm{lim}}$ and $\sigma_{\mathrm{f}}/\bar{\sigma}_{\mathrm{f}}$, with $\bar{K}_{J,\mathrm{lim}}$, and $\bar{\sigma}_{\mathrm{f}}$ representing reference properties. These reference properties are chosen from similar types of existing materials to normalize the mechanical properties obtained. The optimization problem can be summarized as
\begin{equation}
	\max_{\mathbf{x}\in\mathcal{X}} f(\mathbf{x})=
	\max_{\mathbf{x}\in\mathcal{X}} \left[ \lambda\frac{K_{J,\mathrm{lim}}(\mathbf{x})}{\bar{K}_{J,\mathrm{lim}}}+(1-\lambda)\frac{\sigma_{\mathrm{f}}(\mathbf{x})}{\bar{\sigma}_{\mathrm{f}}}\right],
	\label{eq:optimization_problem}
\end{equation}
where $\lambda$ is a weight parameter.

We addressed this problem using BO. Unlike traditional Design of Experiments methods with fixed experiment numbers and limited adaptability, BO follows a sequential approach~\cite{narayanan2021design}. It uses a surrogate model of the actual system and a strategy to decide on the next experiment based on existing data. This helps balance between exploring new areas and exploiting known information~\cite{garnett2023bayesian}.

\subsubsection{Surrogate model}
The Gaussian process (GP) is a probabilistic model extension of the familiar multivariate normal distribution, whose structure can be modified to model functions with a rich variety of behaviors~\cite{williams2006gaussian}. These properties make GP often used to model objective functions in a BO~\cite{garnett2023bayesian}. We denoted the GP representation of $f(\mathbf{x})$ as 
\begin{equation}
	f(\mathbf{x}) \sim \mathcal{GP}(\mu(\mathbf{x}), k(\mathbf{x}, \mathbf{x}')),
	\label{eq:gp}
\end{equation}
where $\mu(\mathbf{x})$ and $k(\mathbf{x}, \mathbf{x}')$ represent the mean function and the covariance function of $f(\mathbf{x})$.
The covariance function plays a critical role in GP as it includes assumptions about the modeled function, and different kernel functions can be used to capture a wide range of patterns and behaviors in the data. We selected the ARD Mat\'ern 5/2 kernel defined as~\cite{snoek2012practical}
\begin{equation}	k_{\mathrm{M52}}(\mathbf{x},\mathbf{x}')=\theta_0(1+\sqrt{5r^2(\mathbf{x},\mathbf{x}')}+\frac{5}{3}r^2(\mathbf{x},\mathbf{x}'))\exp(-\sqrt{5r^2(\mathbf{x},\mathbf{x}')}),
\end{equation}
where $r^2(\mathbf{x},\mathbf{x}')=\sum_{d=1}^{3}(x_d-x_d')^2/\theta^2_d$, and $\{\theta_i\}_{i=0}^3$ are the $i\mathrm{-th}$ kernel hyperparameters.
This kernel exhibits remarkable flexibility in capturing a diverse spectrum of patterns and behaviors in the data, and leads to sample functions that exhibit twice differentiability~\cite{snoek2012practical}.

We can leverage our knowledge of the function $f(\mathbf{x})$ to train the Gaussian Process (GP) and make predictions for its mean value $\mu$ at a given test point $\mathbf{x}_{\ast}$. In practical experiments, we do not have direct access to the true values of $f(\mathbf{x})$, but rather, we have $n$ observations $\{y_i\}_{i=1}^n$ of the noisy function $y=f(\mathbf{x})+\varepsilon$ at specific training points $\{\mathbf{x}_i\}_{i=1}^n$, which are assumed to be precisely known. It is commonly assumed that the noise $\varepsilon$ follows a normal distribution with a mean of zero and a standard deviation of $\sigma_n$. 
The GP problem is fully defined once the hyperparameters $\{\theta_i\}_{i=0}^3$ and $\sigma_n$ are specified. We selected the hyperparameter set that maximize the log marginal likelihood~\cite{williams2006gaussian}
\begin{equation}
	\log p(\mathbf{y}; X) = -\frac{1}{2}\mathbf{y}^{\top}(K(X,X)+\sigma^2_nI)^{-1}\mathbf{y}-\frac{1}{2}\log \left\arrowvert K(X,X)+\sigma^2_nI\right\arrowvert -\frac{n}{2}\log2\pi,
	\label{eq:log_marginal_likehood}
\end{equation}
where $X$ is the $3\times n$ matrix of the training inputs, $I$ is the $n\times n$ identity matrix, $K(X,X)$ is the $n\times n$ covariance matrix where $K_{i,j}=k(\mathbf{x}_i,\mathbf{x}_j)$. 
We calculated the GP posterior mean, $\mu(\mathbf{x}_{\ast})$, and variance, $\sigma^2(\mathbf{x}_{\ast})$, using~\cite{williams2006gaussian}
\begin{equation}
	\label{eq:GPregression_mean}
	\mu(\mathbf{x}_{\ast})= \mathbf{k}^{\mathrm{T}}_{\ast}\left[K(X,X)+\sigma^2_nI\right]^{-1}\mathbf{y},
\end{equation}
\begin{equation}
	\label{eq:GPregression_variance}
	\sigma^2(\mathbf{x}_{\ast})=k(\mathbf{x}_{\ast},\mathbf{x}_{\ast})-\mathbf{k}^{\mathrm{T}}_{\ast}\left[K(X,X)+\sigma^2_nI\right]^{-1} \mathbf{k}_{\ast},
\end{equation}
where $\mathbf{k}_{\ast}$ is the $n\times 1$ covariance matrix between training points and test point.   

To initialize the BO, we trained the GP by selecting an initial set of process parameters based on literature observations (refer to the Meta-experimental optimization methodology section for more detail).

\subsubsection{Acquisition strategy}
The goal of the acquisition strategy is to select the next batch of exploration points for investigation, denoted as $\mathcal{B}_{n_b}=\{\mathbf{x}^b_{1},\dots\mathbf{x}^b_{n_b}\}$, where $n_b$ is the batch size. We adopted the approach introduced by Gonz{\'a}lez et al.~\cite{gonzalez2016batch}, using a sequential algorithm we identified $\mathbf{x}^b_k$ as
\begin{equation}
	\mathbf{x}^b_k=\underset{\mathbf{x}\in\mathcal{X}}{\arg\max}\left[a_k(\mathbf{x};\mathcal{I}_n)\prod_{i=1}^{k-1}\varphi(\mathbf{x};\mathbf{x}^b_i)\right],
\end{equation}
where $a_k(\mathbf{x};\mathcal{I}_n)$ is the $k$-th acquisition function with $\mathcal{I}_n$ representing the available data set $\mathcal{D}_n=\{(\mathbf{x}_i, y_i)\}_{i=1}^n$ and the GP structure, $\varphi(\mathbf{x};\mathbf{x}^b_i)$ are local penalizers centred at $\mathbf{x}^b_i$. We calculated $\varphi(\mathbf{x};\mathbf{x}^b_i)$ as~\cite{gonzalez2016batch}
\begin{equation}
	\varphi(\mathbf{x};\mathbf{x}_i)=\frac{1}{2}\mathrm{erfc}\left[-\frac{1}{\sqrt{2\sigma^2(\mathbf{x}_i)}}(\hat{L}\lVert\mathbf{x}_i-\mathbf{x}\rVert-\hat{M}+\mu(\mathbf{x}_i))\right],
\end{equation}
where $\mathrm{erfc(\cdot)}$ is the complementary error function, $\hat{M}=\max(\{y_i\}_{i=1}^t)$, and we chose a local value of $L$ such that $\hat{L}_i=\lVert \mu_{\nabla}(\mathbf{x}_i)\rVert$, with $\mu_{\nabla}(\mathbf{x}_i)$ the mean of the multivariate Gaussian distribution of $\nabla f(\mathbf{x}_i)$~\cite{gonzalez2016batch}. We chose $a_k(\mathbf{x};\mathcal{I}_n)$ as the expected improvement per unit of cost defined as
\begin{equation}
	a_k(\mathbf{x};\mathcal{I}_n)=\frac{\mathrm{EI}(\mathbf{x};\mathcal{I}_n)}{c_k(\mathbf{x})}.
\end{equation}
where $\mathrm{EI}$ is the expected improvement, and $c_k(\mathbf{x})$ is the cost of each exploration point.
We calculated $\mathrm{EI}$ using~\cite{garnett2023bayesian}
\begin{equation}
	\mathrm{EI}(\mathbf{x};\mathcal{I}_n)=(\mu(\mathbf{x})-y^{\ast}_t) \Phi\left(\frac{\mu(\mathbf{x})-y^{\ast}_t}{\sigma(\mathbf{x})}\right)+\sigma(\mathbf{x})\phi \left(\frac{\mu(\mathbf{x})-y^{\ast}_t}{\sigma(\mathbf{x})}\right),
	\label{eq:expected_improvement}
\end{equation}
where $\Phi(\cdot)$, and $\phi(\cdot)$ are the cumulative distribution function, and the probability density function, respectively, $y^{\ast}$ is the maximum value of $y$ over the set $\mathcal{D}_n$, $\mu(\mathbf{x})$, and $\sigma(\mathbf{x})$ are calculated using equation~\eqref{eq:GPregression_mean}, and equation~\eqref{eq:GPregression_variance}, respectively. We determine the cost function of each exploration point by considering the relative investigation expense for a given exploration point. After selecting the initial exploration point, subsequent points are considered cost-effective for evaluation if the newly chosen process parameters do not involve changes in ceria content or mortar mass ratio. We define the cost of the $k$-th exploration point as
\begin{equation}
	c_k(\mathbf{x})=c_{k-1}+\tilde{c}(\mathbf{x},\mathcal{B}_{k-1}),
	\label{eq:cost}
\end{equation}
where $c_1(\mathbf{x})=1$, and $\tilde{c}(\mathbf{x},\mathcal{B}_{k-1})$ represents the dimensionless evaluation cost. It takes a value of $1$ when the new process parameters $\mathbf{x}$ correspond to an exploration point with the same composition as any point explored in the batch $\mathcal{B}_{k-1}$. Otherwise, it takes a value of $3$. When evaluating the $k$-th point, the cost $c_k$ is a function of the specific point being evaluated. Conversely, the cost $c_{k-1}$ was fixed when the batch point $\mathbf{x}^b_{k-1}$ was selected. Notably, evaluating a point with a different composition is three times costlier than evaluating a point with the same composition as a previously chosen batch point $(\mathbf{x}^b_i\in\mathcal{B}_{k-1})$.

\medskip
\textbf{Acknowledgments} \par 
We gratefully acknowledge the financial support from the Dutch Sectorplan, Zwaartepunt Mechanics of Materials and Zwaartepunt Control Systems Technology. DG gratefully acknowledges the financial support from the Institute of Complex Molecular Systems of TU/e and the Irene Curie Fellowship.

\medskip
\textbf{Conflict of Interest} \par 
The authors declare no conflict of interest.

\bibliographystyle{unsrt}  
\bibliography{references}  

\newpage
\medskip
\textbf{Table of Contents} \par
Double-tough ceramics are new materials that are simultaneously strong and tough. Two toughening mechanisms are embedded, operating at the nano- and microscale. The all-ceramic composition allows no compromise on the ceramics’ key properties: resistance at high-temperature and aggressive environments, strength, biocompatibility, advanced functionalities. Given the broad material design parameter space, Bayesian Optimization is leveraged to smartly guide the experimental campaign.
\begin{figure}[htb]
  \includegraphics[width=55mm]{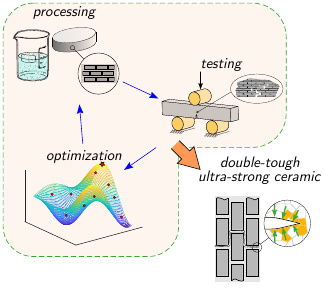}
  \medskip
  \caption*{Schematic representation of the iterative design process and the final all-ceramic composite, with highlighted toughening mechanisms.}
\end{figure}


\newpage

\begin{center}
\textbf{\large Supporting Information}
\end{center}
\textbf{\large Double-tough and ultra-strong ceramics: leveraging multiscale toughening mechanisms through Bayesian Optimization}
\begin{center}
Francesco Aiello,\hspace{5pt}
Jian Zhang,\hspace{5pt}
Johannes C. Brouwer,\hspace{5pt}
Mauro Salazar,\hspace{5pt}
Diletta Giuntini.
\end{center}

\setcounter{equation}{0}
\setcounter{figure}{0}
\setcounter{table}{0}
\setcounter{page}{1}
\renewcommand{\theequation}{S\arabic{equation}}
\renewcommand{\thetable}{S\arabic{table}}
\renewcommand{\thefigure}{S\arabic{figure}}

\begin{figure}[bth]
	\centering
	\includegraphics[width=\textwidth]{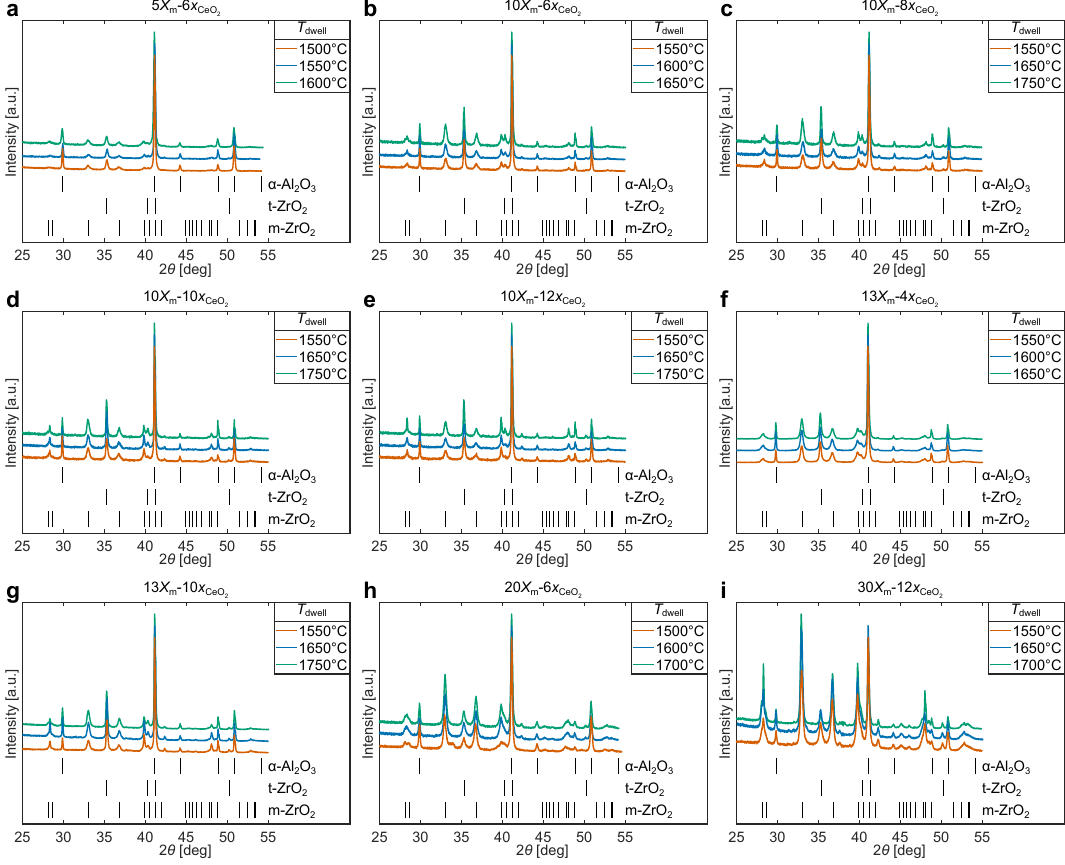}
	\caption{XRD spectra of tested compositions. XRD spectra for different tested combinations of mortar mass ratio ($X_{\mathrm{m}}$), ceria mole fraction ($x_{\mathrm{CeO}_2}$), and dwell temperature ($T_{\mathrm{dwell}}$). Theoretical diffraction peaks for alpha alumina $(\alpha$-Al$_2$O$_3$$)$, tetragonal zirconia $($t-ZrO$_2$$)$, and monoclinic zirconia $($m-ZrO$_2$$)$ are reported in each figure.}
	\label{fig:xrd_spectra}
\end{figure}

Figure~\ref{fig:xrd_spectra}~(a-i) displays XRD spectra for the sintered materials, with the intensity reported on the ordinates relative to the most intense peak. Each figure presents spectra for the same composition at various $T_{\mathrm{dwell}}$. The observed peak positions are close to the theoretical ones of $\alpha$-Al$_2$O$_3$, t-ZrO$_2$, and m-ZrO$_2$ phases, as reported in each figure, and no additional peaks were detected, suggesting the exclusive presence of these three phases under all examined conditions.
\begin{sidewaystable}
\centering
	\caption{Quantities and Crystallite Sizes of $\alpha$-Al$_2$O$_3$, t-ZrO$_2$, and m-ZrO$_2$ phases derived from XRD data refinement tested combinations of mortar mass ratio ($X_{\mathrm{m}}$), ceria mole fraction ($x_{\mathrm{CeO}_2}$), and dwell temperature ($T_{\mathrm{dwell}}$). Standard deviations are reported in brackets.}\label{tbl:xrd_summary}
	\begin{tabular}[htbp]{@{}rrrrrrrrr@{}}
		\toprule
		& & & \multicolumn{6}{c}{Phases} \\ 
		\cmidrule{4-9}
		\multicolumn{3}{c}{Process parameters} & \multicolumn{2}{c}{$\alpha$-Al$_2$O$_3$} & \multicolumn{2}{c}{t-ZrO$_2$} & \multicolumn{2}{c}{m-ZrO$_2$} \\
		\cmidrule{1-3} \cmidrule{4-5} \cmidrule{6-7} \cmidrule{8-9}
		$X_\mathrm{m}$ & $x_{\mathrm{CeO}_2}$ & $T_{\mathrm{dwell}}$ & quantity & cry. size & quantity & cry. size & quantity & cry. size\\
		$(\mathrm{wt.}~\%)$ & $(\mathrm{mol}\%)$ & $(^\circ \mathrm{C})$ & (wt. frac.) & $(\mathrm{nm})$ & (wt. frac.) & $(\mathrm{nm})$ & (wt. frac.) & $(\mathrm{nm})$ \\
		\midrule
		$ 5 $ & $ 6 $ & $ 1500 $ & $ 0.945 ~( 7.5 \!\cdot\! 10^{-4} ) $ & $ 156 ~( 1.9 ) $ & $ 0.023 ~( 4.6 \!\cdot\! 10^{-4} ) $ & $ 54 ~( 1.8 ) $ & $ 0.032 ~( 6.8 \!\cdot\! 10^{-4} ) $ & $ 30 ~( 1.0 ) $ \\
		$ 5 $ & $ 6 $ & $ 1550 $ & $ 0.953 ~( 7.8 \!\cdot\! 10^{-4} ) $ & $ 140 ~( 1.8 ) $ & $ 0.021 ~( 4.6 \!\cdot\! 10^{-4} ) $ & $ 58 ~( 2.4 ) $ & $ 0.027 ~( 7.4 \!\cdot\! 10^{-4} ) $ & $ 29 ~( 1.3 ) $ \\
		$ 5 $ & $ 6 $ & $ 1600 $ & $ 0.945 ~( 8.1 \!\cdot\! 10^{-4} ) $ & $ 97 ~( 1.0 ) $ & $ 0.019 ~( 4.7 \!\cdot\! 10^{-4} ) $ & $ 58 ~( 1.6 ) $ & $ 0.036 ~( 7.9 \!\cdot\! 10^{-4} ) $ & $ 30 ~( 0.9 ) $ \\
		$ 10 $ & $ 6 $ & $ 1550 $ & $ 0.870 ~( 1.5 \!\cdot\! 10^{-3} ) $ & $ 186 ~( 4.3 ) $ & $ 0.054 ~( 9.6 \!\cdot\! 10^{-4} ) $ & $ 80 ~( 2.7 ) $ & $ 0.075 ~( 1.2 \!\cdot\! 10^{-3} ) $ & $ 35 ~( 1.1 ) $ \\
		$ 10 $ & $ 6 $ & $ 1600 $ & $ 0.870 ~( 1.5 \!\cdot\! 10^{-3} ) $ & $ 200 ~( 4.4 ) $ & $ 0.051 ~( 8.7 \!\cdot\! 10^{-4} ) $ & $ 82 ~( 3.0 ) $ & $ 0.080 ~( 1.2 \!\cdot\! 10^{-3} ) $ & $ 36 ~( 0.9 ) $ \\
		$ 10 $ & $ 6 $ & $ 1650 $ & $ 0.852 ~( 1.6 \!\cdot\! 10^{-3} ) $ & $ 156 ~( 3.4 ) $ & $ 0.051 ~( 8.5 \!\cdot\! 10^{-4} ) $ & $ 97 ~( 4.1 ) $ & $ 0.097 ~( 1.4 \!\cdot\! 10^{-3} ) $ & $ 38 ~( 0.9 ) $ \\
		$ 10 $ & $ 8 $ & $ 1550 $ & $ 0.861 ~( 1.5 \!\cdot\! 10^{-3} ) $ & $ 159 ~( 3.2 ) $ & $ 0.058 ~( 9.2 \!\cdot\! 10^{-4} ) $ & $ 68 ~( 2.0 ) $ & $ 0.081 ~( 1.2 \!\cdot\! 10^{-3} ) $ & $ 35 ~( 0.9 ) $ \\
		$ 10 $ & $ 8 $ & $ 1650 $ & $ 0.867 ~( 1.4 \!\cdot\! 10^{-3} ) $ & $ 147 ~( 2.8 ) $ & $ 0.049 ~( 8.3 \!\cdot\! 10^{-4} ) $ & $ 68 ~( 2.2 ) $ & $ 0.084 ~( 1.2 \!\cdot\! 10^{-3} ) $ & $ 36 ~( 0.8 ) $ \\
		$ 10 $ & $ 8 $ & $ 1750 $ & $ 0.850 ~( 1.8 \!\cdot\! 10^{-3} ) $ & $ 130 ~( 3.2 ) $ & $ 0.051 ~( 9.3 \!\cdot\! 10^{-4} ) $ & $ 98 ~( 4.0 ) $ & $ 0.100 ~( 1.5 \!\cdot\! 10^{-3} ) $ & $ 49 ~( 0.9 ) $ \\
		$ 10 $ & $ 10 $ & $ 1550 $ & $ 0.887 ~( 1.4 \!\cdot\! 10^{-3} ) $ & $ 164 ~( 3.4 ) $ & $ 0.048 ~( 8.7 \!\cdot\! 10^{-4} ) $ & $ 61 ~( 2.0 ) $ & $ 0.065 ~( 1.2 \!\cdot\! 10^{-3} ) $ & $ 39 ~( 0.9 ) $ \\
		$ 10 $ & $ 10 $ & $ 1650 $ & $ 0.854 ~( 1.9 \!\cdot\! 10^{-3} ) $ & $ 187 ~( 5.4 ) $ & $ 0.061 ~( 1.1 \!\cdot\! 10^{-3} ) $ & $ 65 ~( 4.1 ) $ & $ 0.084 ~( 1.5 \!\cdot\! 10^{-3} ) $ & $ 39 ~( 1.3 ) $ \\
		$ 10 $ & $ 10 $ & $ 1750 $ & $ 0.845 ~( 2.0 \!\cdot\! 10^{-3} ) $ & $ 206 ~( 6.4 ) $ & $ 0.057 ~( 1.1 \!\cdot\! 10^{-4} ) $ & $ 141 ~( 6.8 ) $ & $ 0.098 ~( 1.7 \!\cdot\! 10^{-3} ) $ & $ 42 ~( 1.2 ) $ \\
		$ 10 $ & $ 12 $ & $ 1550 $ & $ 0.898 ~( 1.5 \!\cdot\! 10^{-3} ) $ & $ 163 ~( 3.8 ) $ & $ 0.046 ~( 1.0 \!\cdot\! 10^{-3} ) $ & $ 64 ~( 2.6 ) $ & $ 0.055 ~( 1.2 \!\cdot\! 10^{-3} ) $ & $ 38 ~( 1.7 ) $ \\
		$ 10 $ & $ 12 $ & $ 1650 $ & $ 0.884 ~( 1.6 \!\cdot\! 10^{-3} ) $ & $ 154 ~( 3.6 ) $ & $ 0.053 ~( 1.1 \!\cdot\! 10^{-3} ) $ & $ 62 ~( 2.3 ) $ & $ 0.064 ~( 1.3 \!\cdot\! 10^{-3} ) $ & $ 39 ~( 1.2 ) $ \\
		$ 10 $ & $ 12 $ & $ 1750 $ & $ 0.862 ~( 2.2 \!\cdot\! 10^{-3} ) $ & $ 174 ~( 5.5 ) $ & $ 0.057 ~( 1.3 \!\cdot\! 10^{-3} ) $ & $ 118 ~( 6.3 ) $ & $ 0.081 ~( 1.8 \!\cdot\! 10^{-3} ) $ & $ 43 ~( 1.7 ) $ \\
		$ 13 $ & $ 4 $ & $ 1550 $ & $ 0.854 ~( 1.3 \!\cdot\! 10^{-3} ) $ & $ 141 ~( 2.6 ) $ & $ 0.047 ~( 7.3 \!\cdot\! 10^{-4} ) $ & $ 49 ~( 1.3 ) $ & $ 0.099 ~( 1.1 \!\cdot\! 10^{-3} ) $ & $ 38 ~( 0.6 ) $ \\
		$ 13 $ & $ 4 $ & $ 1600 $ & $ 0.856 ~( 1.4 \!\cdot\! 10^{-3} ) $ & $ 135 ~( 2.4 ) $ & $ 0.045 ~( 7.5 \!\cdot\! 10^{-4} ) $ & $ 46 ~( 1.2 ) $ & $ 0.098 ~( 1.2 \!\cdot\! 10^{-3} ) $ & $ 37 ~( 0.6 ) $ \\
		$ 13 $ & $ 4 $ & $ 1650 $ & $ 0.853 ~( 1.5 \!\cdot\! 10^{-3} ) $ & $ 130 ~( 2.6 ) $ & $ 0.040 ~( 8.3 \!\cdot\! 10^{-4} ) $ & $ 62 ~( 1.4 ) $ & $ 0.106 ~( 1.3 \!\cdot\! 10^{-3} ) $ & $ 38 ~( 0.7 ) $ \\
		$ 13 $ & $ 10 $ & $ 1550 $ & $ 0.874 ~( 1.6 \!\cdot\! 10^{-3} ) $ & $ 161 ~( 3.9 ) $ & $ 0.057 ~( 1.1 \!\cdot\! 10^{-3} ) $ & $ 80 ~( 2.9 ) $ & $ 0.069 ~( 1.3 \!\cdot\! 10^{-3} ) $ & $ 38 ~( 1.3 ) $ \\
		$ 13 $ & $ 10 $ & $ 1650 $ & $ 0.860 ~( 1.9 \!\cdot\! 10^{-3} ) $ & $ 161 ~( 4.2 ) $ & $ 0.060 ~( 1.1 \!\cdot\! 10^{-3} ) $ & $ 92 ~( 4.0 ) $ & $ 0.080 ~( 1.5 \!\cdot\! 10^{-3} ) $ & $ 40 ~( 1.3 ) $ \\
		$ 13 $ & $ 10 $ & $ 1750 $ & $ 0.860 ~( 1.5 \!\cdot\! 10^{-3} ) $ & $ 174 ~( 3.6 ) $ & $ 0.051 ~( 8.1 \!\cdot\! 10^{-4} ) $ & $ 119 ~( 4.3 ) $ & $ 0.089 ~( 1.2 \!\cdot\! 10^{-3} ) $ & $ 41 ~( 1.0 ) $ \\
		$ 20 $ & $ 6 $ & $ 1500 $ & $ 0.801 ~( 1.6 \!\cdot\! 10^{-3} ) $ & $ 93 ~( 1.3 ) $ & $ 0.035 ~( 8.1 \!\cdot\! 10^{-4} ) $ & $ 34 ~( 0.8 ) $ & $ 0.163 ~( 1.5 \!\cdot\! 10^{-3} ) $ & $ 34 ~( 0.5 ) $ \\
		$ 20 $ & $ 6 $ & $ 1600 $ & $ 0.772 ~( 1.8 \!\cdot\! 10^{-3} ) $ & $ 90 ~( 1.5 ) $ & $ 0.037 ~( 8.5 \!\cdot\! 10^{-4} ) $ & $ 31 ~( 1.1 ) $ & $ 0.190 ~( 1.7 \!\cdot\! 10^{-3} ) $ & $ 35 ~( 0.4 ) $ \\
		$ 20 $ & $ 6 $ & $ 1700 $ & $ 0.767 ~( 1.9 \!\cdot\! 10^{-3} ) $ & $ 88 ~( 1.5 ) $ & $ 0.035 ~( 7.7 \!\cdot\! 10^{-4} ) $ & $ 37 ~( 1.4 ) $ & $ 0.198 ~( 1.8 \!\cdot\! 10^{-3} ) $ & $ 36 ~( 0.5 ) $ \\
		$ 30 $ & $ 12 $ & $ 1550 $ & $ 0.678 ~( 3.6 \!\cdot\! 10^{-3} ) $ & $ 74 ~( 2.1 ) $ & $ 0.047 ~( 1.7 \!\cdot\! 10^{-3} ) $ & $ 35 ~( 1.4 ) $ & $ 0.275 ~( 3.5 \!\cdot\! 10^{-3} ) $ & $ 35 ~( 0.6 ) $ \\
		$ 30 $ & $ 12 $ & $ 1650 $ & $ 0.677 ~( 3.9 \!\cdot\! 10^{-3} ) $ & $ 69 ~( 2.4 ) $ & $ 0.041 ~( 1.6 \!\cdot\! 10^{-3} ) $ & $ 42 ~( 1.8 ) $ & $ 0.282 ~( 3.8 \!\cdot\! 10^{-3} ) $ & $ 40 ~( 0.7 ) $ \\
		$ 30 $ & $ 12 $ & $ 1700 $ & $ 0.675 ~( 6.1 \!\cdot\! 10^{-3} ) $ & $ 68 ~( 4.5 ) $ & $ 0.022 ~( 2.0 \!\cdot\! 10^{-3} ) $ & $ 50 ~( 3.8 ) $ & $ 0.303 ~( 6.2 \!\cdot\! 10^{-3} ) $ & $ 48 ~( 1.1 ) $ \\
		\bottomrule
	\end{tabular}
\end{sidewaystable}
\begin{sidewaystable}
\centering
	\caption{Mechanical properties, mass density, relative density, and objective function values for different tested combinations of mortar mass ratio ($X_{\mathrm{m}}$), ceria mole fraction ($x_{\mathrm{CeO}_2}$), and dwell temperature ($T_{\mathrm{dwell}}$). Measured standard deviations are reported in brackets.}\label{tbl:properties_summary}
	\begin{tabular}[htbp]{@{}rrrrrrrrr@{}}
		\toprule
		\multicolumn{3}{c}{Process parameters} &  \multicolumn{3}{c}{Mechanical properties} & Mass density & Rel. density & Obj. function \\
		\cmidrule{1-3} \cmidrule{4-6} \cmidrule{7-7} \cmidrule{8-8} \cmidrule{9-9}
		$X_\mathrm{m}$ & $x_{\mathrm{CeO}_2}$ & $T_{\mathrm{dwell}}$ & $\sigma_{\mathrm{f}}$ & $K_{J,\mathrm{0}}$ & $K_{J,\mathrm{lim}}$ & $\rho_{\mathrm{s}}$ & $\rho_{\mathrm{r}}$ & $f$ \\
		$(\mathrm{wt.}~\%)$ & $(\mathrm{mol}\%)$ & $(^\circ \mathrm{C})$ & $(\mathrm{MPa})$ & $(\mathrm{MPa}\,\mathrm{m}^{0.5})$ & $(\mathrm{MPa}\,\mathrm{m}^{0.5})$ & $(\mathrm{g\,cm}^{-3})$ & $(\%)$ & $(-)$ \\
		\midrule
		$0$ & $0$ & $1550$ & $487 ~\phantom{0}(44)$ & $5.93 ~(0.33)$ & $10.85 ~(0.10)$ & $3.693$ & $93.1$ & $0.862$ \\
		$0$ & $0$ & $1650$ & $480 ~\phantom{0}(19)$ & $5.24 ~(0.67)$ & $10.10 ~(0.62)$ & $3.830$ & $96.6$ & $0.832$ \\
		$0$ & $0$ & $1750$ & $520 ~\phantom{0}(39)$ & $5.11 ~(0.45)$ & $9.28 ~(0.73)$ & $3.944$ & $99.5$ & $0.852$ \\
		$5$ & $6$ & $1500$ & $289 ~\phantom{0}(23)$ & $4.04 ~(0.64)$ & $7.2\phantom{0}  ~(1.8)\phantom{0}$ & $3.797$ & $94.0$ & $0.533$ \\
		$5$ & $6$ & $1550$ & $508 ~\phantom{0}(53)$ & $4.63 ~(0.45)$ & $8.5\phantom{0}  ~(1.2)\phantom{0}$ & $3.825$ & $95.0$ & $0.815$ \\
		$5$ & $6$ & $1600$ & $542 ~\phantom{0}(23)$ & $3.80 ~(0.19)$ & $7.62 ~(0.18)$ & $3.846$ & $95.2$ & $0.828$ \\
		$10$ & $6$ & $1550$ & $666 ~\phantom{0}(40)$ & $5.71 ~(0.33)$ & $7.8\phantom{0}  ~(1.9)\phantom{0}$ & $4.006$ & $96.6$ & $0.971$ \\
		$10$ & $6$ & $1600$ & $704 ~\phantom{0}(79)$ & $6.70 ~(0.14)$ & $13.64 ~(0.53)$ & $4.114$ & $99.3$ & $1.186$ \\
		$10$ & $6$ & $1650$ & $706~(118)$ & $6.29 ~(0.13)$ & $9.40 ~(0.90)$ & $4.115$ & $98.7$ & $1.063$ \\
		$10$ & $8$ & $1550$ & $523 ~\phantom{0}(38)$ & $5.17 ~(0.36)$ & $5.17 ~(0.36)$ & $4.073$ & $98.0$ & $0.734$ \\
		$10$ & $8$ & $1650$ & $498 ~\phantom{0}(89)$ & $6.22 ~(0.34)$ & $9.8\phantom{0}  ~(1.3)\phantom{0}$ & $4.132$ & $99.6$ & $0.843$ \\
		$10$ & $8$ & $1750$ & $363 ~\phantom{0}(61)$ & $4.42 ~(0.29)$ & $8.64 ~(0.98)$ & $4.113$ & $98.3$ & $0.659$ \\
		$10$ & $10$ & $1550$ & $669 ~\phantom{0}(28)$ & $5.04 ~(0.16)$ & $7.6\phantom{0}  ~(2.2)\phantom{0}$ & $3.976$ & $96.0$ & $0.969$ \\
		$10$ & $10$ & $1650$ & $651~(123)$ & $6.42 ~(0.29)$ & $8.7\phantom{0}  ~(1.3)\phantom{0}$ & $4.083$ & $97.9$ & $0.981$ \\
		$10$ & $10$ & $1750$ & $443 ~\phantom{0}(17)$ & $4.66 ~(0.32)$ & $8.90 ~(0.19)$ & $4.120$ & $98.6$ & $0.756$ \\
		$10$ & $12$ & $1550$ & $580 ~\phantom{0}(39)$ & $5.33 ~(0.08)$ & $6.9\phantom{0}  ~(1.9)\phantom{0}$ & $3.924$ & $95.5$ & $0.848$ \\
		$10$ & $12$ & $1650$ & $537 ~\phantom{0}(64)$ & $4.35 ~(0.24)$ & $7.84 ~(0.49)$ & $4.025$ & $97.7$ & $0.829$ \\
		$10$ & $12$ & $1750$ & $619 ~\phantom{0}(96)$ & $4.61 ~(0.43)$ & $8.33 ~(0.70)$ & $4.100$ & $98.7$ & $0.934$ \\
		$13$ & $4$ & $1550$ & $682 ~\phantom{0}(49)$ & $6.45 ~(0.56)$ & $7.68 ~(0.42)$ & $4.053$ & $97.3$ & $0.985$ \\
		$13$ & $4$ & $1600$ & $710 ~\phantom{0}(15)$ & $6.21 ~(0.34)$ & $8.85 ~(2.9)\phantom{0}$ & $4.054$ & $97.3$ & $1.051$ \\
		$13$ & $4$ & $1650$ & $672 ~\phantom{0}(29)$ & $5.94 ~(0.42)$ & $7.79 ~(1.6)\phantom{0}$ & $4.098$ & $98.3$ & $0.977$ \\
		$13$ & $10$ & $1550$ & $541 ~\phantom{0}(45)$ & $4.16 ~(0.20)$ & $4.24 ~(0.23)$ & $4.133$ & $99.9$ & $0.726$ \\
		$13$ & $10$ & $1650$ & $514 ~\phantom{0}(61)$ & $3.81 ~(0.53)$ & $5.96 ~(0.64)$ & $4.142$ & $99.6$ & $0.748$ \\
		$13$ & $10$ & $1750$ & $433 ~\phantom{0}(84)$ & $3.46 ~(0.67)$ & $6.07 ~(0.91)$ & $4.144$ & $99.7$ & $0.661$ \\
		$20$ & $6$ & $1500$ & $226 ~\phantom{0}(35)$ & $3.53 ~(0.20)$ & $6.33 ~(0.54)$ & $3.968$ & $93.5$ & $0.438$ \\
		$20$ & $6$ & $1600$ & $569 ~\phantom{0}(40)$ & $5.97 ~(0.50)$ & $10.8\phantom{0}  ~(1.2)\phantom{0}$ & $4.220$ & $98.5$ & $0.952$ \\
		$20$ & $6$ & $1700$ & $526 ~\phantom{0}(24)$ & $5.85 ~(0.32)$ & $10.6\phantom{0}  ~(1.0)\phantom{0}$ & $4.247$ & $99.0$ & $0.898$ \\
		$30$ & $12$ & $1550$ & $575~(125)$ & $4.79 ~(0.11)$ & $8.79 ~(0.04)$ & $4.399$ & $99.3$ & $0.899$ \\
		$30$ & $12$ & $1650$ & $567 ~\phantom{0}(80)$ & $5.45 ~(0.52)$ & $9.28 ~(0.35)$ & $4.397$ & $99.3$ & $0.905$ \\
		$30$ & $12$ & $1700$ & $311 ~\phantom{0}(67)$ & $3.90 ~(0.16)$ & $7.10 ~(0.34)$ & $4.364$ & $98.5$ & $0.555$ \\
		\bottomrule
	\end{tabular}
\end{sidewaystable}

\end{document}